\documentclass[aip,pop,reprint]{revtex4-1}
\usepackage{graphicx}
\usepackage{amsmath, amssymb, mathtools}
\usepackage[dvipsnames]{xcolor}
\usepackage{epstopdf, epsfig}
\usepackage{hyperref, cleveref}
\usepackage{comment}
\usepackage{dcolumn}
\usepackage{bm}
\usepackage{longtable}

\usepackage[utf8]{inputenc}
\usepackage[T1]{fontenc}
\usepackage{mathptmx}

\allowdisplaybreaks

\makeatletter
\def\@cline#1-#2\@nil{%
  \omit
  \@multicnt#1%
  \advance\@multispan\m@ne
  \ifnum\@multicnt=\@ne\@firstofone{&\omit}\fi
  \@multicnt#2%
  \advance\@multicnt-#1%
  \advance\@multispan\@ne
  \leaders\hrule\@height\arrayrulewidth\hfill
  \cr
  \noalign{\nobreak\vskip-\arrayrulewidth}}
\makeatother

\draft

\begin{document}

\title[A two-fluid analysis of waves in a warm ion-electron plasma]{A two-fluid analysis of waves in a warm ion-electron plasma}

\author{J. De Jonghe}
\thanks{ORCID 0000-0003-2443-3903}
\email{jordi.dejonghe@kuleuven.be}
\affiliation{Centre for mathematical Plasma-Astrophysics, KU Leuven, 3001 Leuven, Belgium.}

\author{R. Keppens}
\thanks{ORCID 0000-0003-3544-2733}
\affiliation{Centre for mathematical Plasma-Astrophysics, KU Leuven, 3001 Leuven, Belgium.}

\date{\today}

\begin{abstract}
Following recent work, we discuss waves in a warm ideal two-fluid plasma consisting of electrons and ions starting from a completely general, ideal two-fluid dispersion relation. The plasma is characterised by five variables: the electron and ion magnetisations, the squared electron and ion sound speeds, and a parameter describing the angle between the propagation vector and the magnetic field. The dispersion relation describes 6 pairs of waves which we label S, A, F, M, O, and X. Varying the angle, it is argued that parallel and perpendicular propagation (with respect to the magnetic field) exhibit unique behaviour. This behaviour is characterised by the crossing of wave modes which is prohibited at oblique angles. We identify up to 6 different parameter regimes where a varying number of exact mode crossings in the special parallel or perpendicular orientations can occur. We point out how any ion-electron plasma has a critical magnetisation (or electron cyclotron frequency) at which the cutoff ordering changes, leading to different crossing behaviour. These are relevant for exotic plasma conditions found in pulsar and magnetar environments. Our discussion is fully consistent with ideal relativistic MHD and contains light waves. Additionally, exploiting the general nature of the dispersion relation, phase and group speed diagrams can be computed at arbitrary wavelengths for any parameter regime. Finally, we recover earlier approximate dispersion relations that focus on low-frequency limits and make direct correspondences with some selected kinetic theory results.
\end{abstract}

\maketitle

\section{Introduction}
In many textbooks wave propagation is discussed at length for plasmas consisting of ions and electrons \cite{Stix1992, BoydSanderson2003, GurnettBhattacharjee2005, ThorneBlandford2017, GoedbloedKeppensPoedts2019}. There are many approaches to this topic and often these textbooks opt for a two-fluid approach as a way to introduce different wave types. However, in the literature this approach is not explored to its full extent. In particular, textbooks usually discuss waves at parallel and perpendicular propagation with respect to the background magnetic field even though these are limit cases and not representative of wave behaviour at oblique angles. In this paper we will point out the differences between parallel, perpendicular, and oblique propagation using the S, A, F, M, O, X wave labelling scheme introduced in Refs. \onlinecite{Keppens2019_coldpair, Keppens2019_coldei, Keppens2019_warmpair}. In fact, this special status of parallel and perpendicular propagation was already noted in Ref. \onlinecite{Stringer1963} for low-frequency waves. In particular, it was argued that if two modes at parallel propagation can exist with identical frequency and wavenumber, it is natural to interpret this frequency-wavenumber pair as a crossing of these modes in the frequency-wavenumber diagram. This transition from mode crossings at parallel propagation to avoided crossings at oblique angles was also the subject of the more recent work in Refs. \onlinecite{Keppens2019_coldpair, Keppens2019_coldei, Keppens2019_warmpair}. In this paper a similar analysis is performed for a warm ion-electron plasma.

The two-fluid treatment here, described in Ref. \onlinecite{GoedbloedKeppensPoedts2019}, employs a polynomial description of sixth order in the squared frequency ($\omega^2$), thus describing six wave types, and goes back to the description of Ref. \onlinecite{DenisseDelcroix1961}. In up to recent literature however, similar approaches have been used to focus solely on low-frequency waves \cite{Stringer1963, Ishida2005, Damiano2009, Bellan2012, Zhao2014, Zhao2015}, related to the MHD slow, Alfv\'en, and fast modes, medium-frequency waves \cite{Huang2019} or high-frequency waves, e.g. in the form of the Appleton-Hartree description \cite{Appleton1932}. Since the fully general dispersion relation was used successfully for pair plasmas \cite{Keppens2019_coldpair, Keppens2019_warmpair} and cold ion-electron plasmas \cite{Keppens2019_coldei}, and thereby revealed surprising new insights (such as on how the Appleton-Hartree limit is deduced as an unphysical limit in practice), we use this sixth order dispersion relation to expand this study to warm ion-electron plasmas.

Starting from a homogeneous background consisting of ions and electrons at rest subject to a uniform magnetic field $\mathbf{B}$, small amplitude oscillations are inserted and plane wave solutions ($\sim\exp[i(\mathbf{k}\cdot\mathbf{x}-\omega t)]$) are assumed. The plasma considered here is charge-neutral, i.e. $n_\mathrm{e} = Zn_\mathrm{i}$, where $n_\mathrm{e}$ and $n_\mathrm{i}$ denote the electron and ion number densities, and $Z$ is the ion charge number. Each species introduces a continuity equation, a momentum equation, and an energy equation. These are complemented by Maxwell's equations. After linearisation the dispersion relation is obtained as a function of the wave frequency $\omega$ and wavenumber $k = |\mathbf{k}|$. The dispersion relation is a polynomial in $\omega^2$ and $k^2$ because all waves come in forward and backward propagating pairs. A detailed derivation can be found in Ref. \onlinecite{GoedbloedKeppensPoedts2019}.

The coefficients in this polynomial are determined by various physical quantities of the plasma. These quantities are defined separately for each species, namely the electron and ion plasma frequencies $\omega_\mathrm{pe} = \sqrt{e^2n_\mathrm{e}/\epsilon_0m_\mathrm{e}}$ and $\omega_\mathrm{pi} = \sqrt{Z^2e^2n_\mathrm{i}/\epsilon_0m_\mathrm{i}}$, cyclotron frequencies $\Omega_\mathrm{e} = eB/m_\mathrm{e}$ and $\Omega_\mathrm{i} = ZeB/m_\mathrm{i}$, and sound speeds $v_\mathrm{e} = \sqrt{\gamma p_\mathrm{e}/n_\mathrm{e} m_\mathrm{e}}$ and $v_\mathrm{i} = \sqrt{\gamma p_\mathrm{i}/n_\mathrm{i} m_\mathrm{i}}$. Here, we denote the strength of the background uniform magnetic field with $B$ and adopt a ratio of specific heats $\gamma$. For each species $s$, $m_s$ and $p_s$ denote the mass and pressure, respectively. The plasma frequency is then given by $\omega_\mathrm{p} = \sqrt{\omega_\mathrm{pe}^2+\omega_\mathrm{pi}^2}$. Throughout this paper these quantities will only appear in a dimensionless way, adopting the notation of Ref. \onlinecite{GoedbloedKeppensPoedts2019},
\begin{equation}
\begin{aligned}
e &\equiv \omega_\mathrm{pe}/\omega_\mathrm{p}, \qquad &&E \equiv \Omega_\mathrm{e}/\omega_\mathrm{p}, \qquad &&&v \equiv v_\mathrm{e}/c, \\
i &\equiv \omega_\mathrm{pi}/\omega_\mathrm{p}, \qquad &&I \equiv \Omega_\mathrm{i}/\omega_\mathrm{p}, \qquad &&&w \equiv v_\mathrm{i}/c.
\end{aligned}
\end{equation}
Introducing the ratio of masses over charges $\mu = Zm_\mathrm{e}/m_\mathrm{i}$, these satisfy the relations
\begin{equation}
e^2 = \frac{1}{1+\mu}, \qquad i^2 = \frac{\mu}{1+\mu}, \qquad I = \mu E.
\end{equation}
The final parameter is defined as $\lambda = \cos\theta$ where $\theta$ is the angle between the wavevector $\mathbf{k}$ and the magnetic field $\mathbf{B}$. This results in a set of five dimensionless parameters $E$, $\mu$, $v$, $w$, and $\lambda$ to characterise all wave types.

To write the entire dispersion relation in dimensionless quantities the frequency and wavenumber are also normalised as
\begin{equation}
\bar{\omega} = \omega/\omega_\mathrm{p}, \qquad \bar{k} = ck/\omega_\mathrm{p}
\end{equation}
where $c/\omega_\mathrm{p} \equiv \delta$ is the combined skin depth. The result is a polynomial of sixth order in $\bar{\omega}^2$ describing six wave types (S, A, F, M, O, and X). Here, the first three types are related to the Alfv\'en (A) and slow (S) and fast (F) magnetoacoustic modes in the low-frequency, long wavelength MHD limit. The remaining three types are the modified electrostatic wave (M) and two electromagnetic waves (O and X).

To lighten the notation in the remainder of the paper, all bars are omitted but implied unless specified otherwise.

\section{Dispersion relation}\label{sec:disprel}
The dispersion relation of a warm ion-electron plasma is a polynomial of sixth order in $\omega^2$ and fourth order in $k^2$. Furthermore, the combined order of $\omega^2$ and $k^2$ is at least three and at most six. The dispersion relation can thus be written as \cite{GoedbloedKeppensPoedts2019}
\begin{equation}\label{eq:disp-rel}
\sum\limits_{\substack{0 \leq m \leq 6 \\ 0 \leq n \leq 4 \\ 3 \leq m+n \leq 6}} \alpha_{mn}\, \omega^{2m} k^{2n} = 0.
\end{equation}
The coefficient notation $\alpha_{mn}$ will be used to discuss particular terms in the rest of this paper and follows the conventions of Ref. \onlinecite{GoedbloedKeppensPoedts2019}. The coefficient expressions themselves can be found in App. \ref{app:disprel}. Note that the expression $i^2 v^2 + e^2 w^2$ here has been related to the normalised, combined ion-electron sound speed,
\begin{equation}\label{eq:sound}
c_\mathrm{s}^2 \equiv \frac{v_\mathrm{s}^2}{c^2} = i^2 v^2 + e^2 w^2.
\end{equation}

Denoting the dispersion relation in this way has several distinct advantages not the least of which is the relative ease with which we can extract limit behaviour. Various limits were already indicated in Ref. \onlinecite{GoedbloedKeppensPoedts2019}. The global, low-frequency limit encompassing the terms of combined order three reduces to MHD whilst the terms of combined order six offer the local, high-frequency behaviour. The terms of zeroth order in $k^2$ ($\alpha_{m0}$) offer the cutoff limit and those of fourth order in $k^2$ ($\alpha_{m4}$) the resonance limit.

To label the different wave types and study their behaviour it is necessary to examine their long and short wavelength limits as well as how these regimes connect. To answer this question, we look for (avoided) crossings of modes, i.e. pairs of $(\omega, k)$ that correspond to multiple modes. A first step towards finding such crossings consists of checking when the dispersion relation factorises. Unfortunately, the dispersion relation does not factorise at all angles like it does in the case of a warm pair plasma \cite{Keppens2019_warmpair}. However, the dispersion relation does factorise in the edge cases of parallel and perpendicular propagation with respect to the magnetic field ($\lambda = 1$ and $\lambda = 0$ respectively). As we will show, this is indicative of a transition between crossings and avoided crossings.

Now, before discussing limit behaviour of the dispersion relation, this section will first introduce the labelling scheme unambiguously. Next, we establish a basis of comparison by going over the results of an unmagnetised ion-electron plasma. Then, the limit behaviour of the dispersion relation with respect to the frequency and the wavenumber is treated. Finally, the factorisation of the dispersion relation is shown for parallel and perpendicular propagation. How the limits discussed in Sec. \ref{sec:limitbehaviour} appear in the factorised parallel and perpendicular dispersion relations is the subject of Sec. \ref{sec:dispdiagrams}, where they are presented visually.

\subsection{Wave labels}\label{sec:labels}
Before labelling waves, we first motivate our choice of labels. The wave labels A, F, M, O, and X were originally introduced in Ref. \onlinecite{Keppens2019_coldpair} and the S label followed in the warm pair plasma case \cite{Keppens2019_warmpair}. As we will show, at oblique angles the frequency ordering of the modes is fixed in the order
\begin{equation*}
\omega_\mathrm{S} \leq \omega_\mathrm{A} \leq \omega_\mathrm{F} \leq \omega_\mathrm{M} \leq \omega_\mathrm{O} \leq \omega_\mathrm{X},
\end{equation*}
which allows for consistent labelling. However, at parallel and perpendicular propagation these modes cross. Therefore, it is important to motivate how the labels are assigned at all angles.

The S, A, and F labels are used for the waves that behave like the well-established MHD Alfv\'en (A), and slow (S) and fast (F) magnetoacoustic waves in the long wavelength (small $k$) limit. Hence, at any angle these modes are labelled by their \emph{long wavelength behaviour} such that the MHD ordering $\omega_\mathrm{S} \leq \omega_\mathrm{A} \leq \omega_\mathrm{F}$ is satisfied near $k = 0$.

The M label stands for modified electrostatic mode and corresponds at parallel propagation to the textbook Langmuir wave, which is given by the dispersion relation
\begin{equation}\label{eq:langmuir}
\omega^2 = \omega_\mathrm{pe}^2 + \gamma_\mathrm{e} C_\mathrm{e}^2 k^2 \qquad \text{(no bars implied)}.
\end{equation}
Here, the second term comes from the electron pressure \cite{GurnettBhattacharjee2005}. Note that we temporarily distinguish a ratio of specific heats for electrons in $\gamma_\mathrm{e}$. In our conventions, eq. (\ref{eq:langmuir}) implies short wavelength behaviour (large $k$) of $\omega^2 \simeq k^2 v^2$. Therefore, we identify the M mode with this \emph{short wavelength behaviour} at all angles. Finally, the O and X labels are borrowed from the literature which speaks of ordinary (O) and extraordinary (X) electromagnetic modes. The short wavelength behaviour of either mode is electromagnetic, meaning $\omega = k$, but they are distinguished from each other by their long wavelength behaviour where they have a different cutoff value $\omega^2(k^2\rightarrow 0) = \text{cst} > 0$. The X mode has the highest cutoff value of all wave types whereas the O mode has the lower cutoff of these two modes.

It should be pointed out that this convention of labelling the wave modes differs for the M, O, and X modes from the conventions adopted in Refs. \onlinecite{Keppens2019_coldpair, Keppens2019_coldei, Keppens2019_warmpair}. There, the M, O, and X modes were also labelled by imposing the ordering $\omega_\mathrm{M} \leq \omega_\mathrm{O} \leq \omega_\mathrm{X}$, but this was done for $k = 0$. This leads to undesirable ambiguity in the warm ion-electron case (see Sec. \ref{sec:cutoff-limit}) which is the reason for this deviation from previous works.

As a final note, this paper offers various graphical representations of the different modes. Throughout the whole paper, a consistent colour convention is adopted identical to the one in Ref. \onlinecite{Keppens2019_warmpair}. Each mode is always represented by its own colour: S$-$green, A$-$red, F$-$blue, M$-$purple, O$-$cyan, and X$-$black.

\subsection{Unmagnetised warm ion-electron plasma}\label{sec:unmagnetised}
With our definiton of the wave labels we first consider the unmagnetised case to establish a basis of comparison. For an unmagnetised plasma ($B = 0$), the parameters $E$ and $I$ are zero, which naturally eliminates all terms featuring $\lambda$. However, $\mu = I/E$ remains a finite nonzero constant. The remaining three parameters describing all waves are thus the electron and ion sound speeds $v$ and $w$, and the ratio of masses over charges $\mu$. The factorised dispersion relation becomes
\begin{equation}\label{eq:disp-unmagnetised}
\begin{aligned}
&\omega^{4} \left(\omega^{2} - k^2 - 1\right)^{2} \\
\times &\{ \omega^{4} - \omega^{2} [1 + k^2 (v^2 + w^2)] + k^2 [c_\mathrm{s}^2 + k^2 v^{2} w^{2}] \} = 0.
\end{aligned}
\end{equation}
There are two degenerate first order branches and one quadratic branch in $\omega^2$. As could be anticipated from MHD, two trivial solutions $\omega^2 = 0$ describe the S (slow) and A (Alfv\'en) modes which do not propagate in the absence of a magnetic field. To label the remaining modes we consider the small and large wavenumber limits of the dispersion relation (\ref{eq:disp-unmagnetised}). In the small wavenumber limit the quadratic branch reduces to $0 = \omega^2(\omega^2-1)$ whereas the other two modes reduce to $\omega^2 = 1$. Hence, we can already conclude that the quadratic branch describes the F mode thanks to the $\omega^2$ factor. Furthermore, the quadratic branch will also describe either the M, O, or X mode. To ascertain which one it is, we consider the local, high-frequency limit ($\omega^2 \rightarrow\infty$, $k^2\rightarrow\infty$, $\omega^2/k^2$ finite) of the quadratic branch. In this limit the dispersion relation can be written as
\begin{equation}
\left( \omega^2 - k^2 v^2 \right) \left( \omega^2 - k^2 w^2 \right) = 0.
\end{equation}
The first factor in this expression describes behaviour of the form $\omega^2 = k^2 v^2$ which we defined to be the M mode. Thus, we conclude that the quadratic branch describes the F and M mode. The second factor in the expression above then gives the short wavelength behaviour of the F mode.

The global, low-frequency limit ($\omega^2\rightarrow 0$, $k^2\rightarrow 0$, $\omega^2/k^2$ finite) retrieves the long wavelength behaviour of the acoustic F mode, $\omega^2 = k^2 c_\mathrm{s}^2$, which is how eq. (\ref{eq:sound}) was obtained. Note that $\min\{v, w\} < c_\mathrm{s} < \max\{v, w\}$. From now on we will assume that $v > w$ such that the ordering becomes $w < c_\mathrm{s} < v$. This assumption of having an electron thermal velocity above the ion thermal velocity is fairly representative of expected behaviour where ions are more immobile. A plasma where $w > v$ can be analysed as well from our general dispersion relation, but is outside the scope of this paper. In the limit case $v = w$, and thus $c_\mathrm{s} = v = w$, the quadratic branch factorises further into
\begin{equation}
(\omega^2 - k^2c_\mathrm{s}^2) (\omega^2 - 1 - k^2c_\mathrm{s}^2) = 0
\end{equation}
and both modes have the same short wavelength (large $k$) behaviour.

\subsection{Cutoffs and resonances}\label{sec:limitbehaviour}
Before turning to the factorisation of the dispersion relation, it is instructive to discuss the limit behaviour of that dispersion relation. In particular, the long or short wavelength behaviour of a mode is used to label it. Additionally, how each mode connects to the other regime is one of the key questions we wish to answer. The cutoff, resonance and local, high-frequency limit were already offered in Ref. \onlinecite{GoedbloedKeppensPoedts2019} whereas the global, low-frequency limit was only given in approximate form. Since they are invaluable to our discussion though, they are reproduced and discussed here.

\subsubsection{Cutoffs}\label{sec:cutoff-limit}
Starting with the cutoff limit the dispersion relation gives
\begin{equation}\label{eq:cutoffs}
\omega^2(k^2 \rightarrow 0) = \left\{ \begin{aligned}
&1, \\
&1 + \frac{1}{2}(E^2+I^2) \pm \frac{1}{2} |E-I| \sqrt{(E+I)^2+4} \\ &\qquad\equiv \omega_\mathrm{u,l}^2,
\end{aligned} \right.
\end{equation}
where $\omega_\mathrm{u}$ signifies the plus (upper) sign and $\omega_\mathrm{l}$ the minus (lower) sign. First of all, note that the triple $\omega^2 = 1$ degeneracy in the cutoff limit of the unmagnetised case has been lifted by applying a magnetic field. Secondly, a pair plasma satisfies $E = I$ and substituting this in the equation recovers the result ($5.1$) from Ref. \onlinecite{Keppens2019_warmpair}. Thirdly, note that $\omega_\mathrm{u}$ is always larger than $1$ and $\omega_\mathrm{l}$. Therefore, it will always correspond to the X mode. Additionally, using $I = \mu E$ it can be shown that $\omega_\mathrm{l} < 1$ if and only if
\begin{equation}\label{eq:E-crit}
E < \frac{1-\mu}{\mu} \qquad\equiv E_{\mathrm{cr}}.
\end{equation}
This was already briefly pointed out in Ref. \onlinecite{Keppens2019_coldei}, but will be discussed in more detail in the present paper. We will refer to this value $E_{\mathrm{cr}}$ as the critical electron cyclotron frequency or critical magnetisation. Alternatively, this critical value can be expressed in terms of the electron and ion plasma frequencies, $E_{\mathrm{cr}} = (\omega_{\mathrm{pe}}^2-\omega_{\mathrm{pi}}^2)/\omega_{\mathrm{pi}}^2$, or we can define a critical (normalised) Alfv\'en speed, $c_{\mathrm{a,cr}} = (1-\mu)/\sqrt{1-\mu+\mu^2}$. This result is the main reason why the M mode is defined by its short wavelength behaviour rather than by the ordering at $k^2 = 0$. This definition of the M mode (and consequently the O mode) results in a consistent labelling across this critical value.

As it turns out, this quantity $E_{\mathrm{cr}}$ will also appear multiple times in the discussion to follow. It is also worth pointing out that for a pair plasma ($\mu = 1$) the critical electron cyclotron frequency is zero such that any magnetised pair plasma satisfies $\omega_\mathrm{l} > 1$. This is confirmed by the result ($5.1$) in Ref. \onlinecite{Keppens2019_warmpair} stating $\omega_\mathrm{u,l}^2 = 1+E^2$.

Since $\omega_\mathrm{l} = 1$ for $E = 0$ and $\omega_\mathrm{l} < 1$ for $0 < E < E_{\mathrm{cr}}$ there must also be a value of $E$ for which $\omega_\mathrm{l}$ is minimal. This minimising value of $E$ can be computed to be
\begin{equation}
E_{\mathrm{m}} = \frac{1}{\sqrt{\mu}} \frac{1-\mu}{1+\mu}.
\end{equation}
In agreement with our previous remark on the result obtained in Ref. \onlinecite{Keppens2019_warmpair}, the minimising value of $E$ is zero for a pair plasma ($\mu=1$).

\subsubsection{Resonances}\label{sec:resonance-limit}
Two resonances are present,
\begin{equation}\label{eq:resonances}
\omega^2(k^2 \rightarrow \infty) = \left\{ \begin{aligned}
&\lambda^2 E^2,\\
&\lambda^2 I^2.
\end{aligned} \right.
\end{equation}
Due to the $\lambda$ dependence there is a clear difference between parallel and perpendicular propagation. In the case of parallel ($\lambda = 1$) and oblique propagation ($0 < \lambda < 1$) two modes will display a resonance, the ion and electron cyclotron resonances. For perpendicular propagation ($\lambda = 0$) both resonances are absent.

\subsubsection{Local, high-frequency limit}\label{sec:lhf-limit}
Considering there are two resonances, the other four modes are expected to behave linearly for $k^2\rightarrow\infty$. Taking the local, high-frequency limit ($\omega^2\rightarrow\infty, k^2 \rightarrow \infty, \omega^2/k^2$ finite) confirms this expectation and gives
\begin{equation}\label{eq:local-high-f}
\frac{\omega^2}{k^2} \rightarrow \left\{ \begin{aligned}
&1 \quad\text{(twice)},\\
&v^2, \\
&w^2.
\end{aligned} \right.
\end{equation}
Two modes show electromagnetic behaviour and the remaining two behave acoustically with the ion and electron sound speeds respectively. Further note that in the case of a pair plasma the result ($5.3$) from Ref. \onlinecite{Keppens2019_warmpair} is obtained by using equal sound speeds, $v = w$.

\subsubsection{Global, low-frequency limit}\label{sec:glf-limit}
Finally, examining the global, low-frequency (MHD) limit ($\omega^2, k^2 \rightarrow 0, \omega^2/k^2$ finite) produces
\begin{equation}\label{eq:global-low-f}
\frac{\omega^2}{k^2} \rightarrow \left\{ \begin{aligned}
&\frac{\lambda^2 EI}{1+EI},\\
&v_\mathrm{sl,f}^2 \equiv \frac{1}{2(1 + EI)} \bigg\{ EI + c_\mathrm{s}^2 + \lambda^2 EI c_\mathrm{s}^2 \pm \bigg[ \lambda^4 E^2 I^2 c_\mathrm{s}^4 \\ &\hspace{1cm} + 2\lambda^2 EI c_\mathrm{s}^2 (c_\mathrm{s}^2-EI-2) + (EI+c_\mathrm{s}^2)^2 \bigg]^{1/2} \bigg\}
\end{aligned} \right.
\end{equation}
The global, low-frequency limit can be reduced to MHD so the three modes described by this behaviour are related to the MHD slow, Alfv\'en, and fast waves. The first result describes the Alfv\'en wave with the relativistic expression for the normalised squared Alfv\'en speed, $c_\mathrm{a}^2 \equiv v_\mathrm{a}^2/c^2 = EI / (1+EI)$. (To avoid any possible confusion with the A mode, we use a lowercase `a' subscript to denote the Alfv\'en speed.)

The second expression corresponds to the fast ($+$) and slow ($-$) waves. Reorganising the expression and comparing to eq. ($22.130$) in Ref. \onlinecite{GoedbloedKeppensPoedts2019} shows that we recover the relativistic phase speed for the fast and slow MHD waves. Furthermore, it can also be compared to the cold ion-electron plasma case and the warm pair plasma case like before. Setting the sound speed $c_\mathrm{s}$ to zero reduces the fast speed $v_\mathrm{f}$ to the Alfv\'en speed in agreement with eq. ($16$) in Ref. \onlinecite{Keppens2019_coldei}. The slow speed $v_\mathrm{sl}$ vanishes for a cold setting. For a warm pair plasma this expression reduces to eq. ($4.7$) in Ref. \onlinecite{Keppens2019_warmpair} by substituting $c_\mathrm{s}^2 = v^2$ and $I = E$.

Moreover, the expression of $v_\mathrm{sl,f}^2$ simplifies significantly in the cases of parallel and perpendicular propagation. For perpendicular propagation ($\lambda = 0$) the expression reduces to $v_{\perp,\mathrm{f}}^2 = (EI+c_\mathrm{s}^2)/(1+EI)$ and $v_{\perp,\mathrm{sl}}^2 = 0$. As expected, only the fast wave propagates with a behaviour of $\omega^2 = k^2 v_{\perp,\mathrm{f}}^2$.

For parallel propagation it is a bit more nuanced. The square root in eq. (\ref{eq:global-low-f}) simplifies to $|c_\mathrm{s}^2 (1+EI) - EI|$. Therefore, at parallel propagation we get $v_{\parallel,\mathrm{f}}^2 = \max\{ c_\mathrm{a}^2, c_\mathrm{s}^2 \}$ and $v_{\parallel,\mathrm{sl}}^2 = \min\{ c_\mathrm{a}^2, c_\mathrm{s}^2 \}$ in accordance with MHD. The low-frequency limit expressions are ordered as $v_\mathrm{sl} \leq \lambda c_\mathrm{a} \leq v_\mathrm{f}$ and this corresponds to the ordering of the slow (S), Alfv\'en (A), and fast (F) frequencies $\omega_\mathrm{S} \leq \omega_\mathrm{A} \leq \omega_\mathrm{F}$.

Now that the relevant limits have been identified, it can be studied how they appear in the different branches of parallel and perpendicular propagation. The interest lies in how the modes connect the long wavelength limits to the short wavelength limits and the crossings or avoided crossings that appear.

\subsection{Parallel propagation}\label{sec:disprel-pl}
In the case of parallel propagation the parameter $\lambda = 1$ is fixed such that there are four remaining parameters ($E$, $I$, $v$, and $w$). The dispersion relation splits into a quadratic branch,
\begin{equation}\label{eq:branch-pl2}
0 = \omega^{4} - \omega^{2} [1 + k^2 (v^2 + w^2)] + k^{2} (c_\mathrm{s}^2 + k^{2} v^{2} w^{2}) ,
\end{equation}
and a quartic branch,
\begin{equation}\label{eq:branch-pl4}
\begin{aligned}
&\omega^{8} - \omega^{6} \left[ 2 + E^2 + I^2 + 2 k^{2} \right] && \\
&+ \omega^{4} \left[ \left(1 + EI \right)^2 + 2 k^{2} \left( 1 + E^2 + I^2 \right) + k^{4} \right] && \\ &- \omega^{2}k^2 \left[ 2 EI \left(1 + EI \right) + k^{2} (E^2+I^2) \right] + k^{4} E^2 I^2 &&= 0.
\end{aligned}
\end{equation}
First of all, note that the quadratic branch is the same expression as the one found in eq. (\ref{eq:disp-unmagnetised}) of the unmagnetised case. However, labelling the waves described by this branch is a little less straightforward this time. As shown in Sec. \ref{sec:unmagnetised}, this quadratic branch has one mode that behaves like $\omega^2 = k^2 v^2$ in the short wavelength limit. This is undisputably the M mode. The label of the second mode on the other hand depends on the relation between the sound speed $c_\mathrm{s}$ and the Alfv\'en speed $c_\mathrm{a}$. The long wavelength limit of this mode goes like $\omega^2 = k^2 c_\mathrm{s}^2$ (see Sec. \ref{sec:unmagnetised}) which corresponds to the S mode if $c_\mathrm{s} < c_\mathrm{a}$ or to the F mode if $c_\mathrm{s} > c_\mathrm{a}$. Naturally, this implies that the quartic branch describes the A, O, and X modes as well as the remaining mode of the S or F variety.

Secondly, looking at the expressions of both branches, it is immediately clear that only the quadratic branch features the sound speeds $v$ and $w$ (and the linear combination $c_\mathrm{s}$ thereof) whereas only the quartic branch is influenced by the magnetic field ($E$, $I$). Thus, reducing these expressions to the cold case by imposing $v = w = c_\mathrm{s} = 0$ does not alter the quartic branch, but does reduce the quadratic branch to $\omega^2 (\omega^2-1)$. In doing so, we effectively recover the cold plasma result of Ref. \onlinecite{Keppens2019_coldei} after decomposing the quartic branch into two fourth order factors in $\omega$ rather than $\omega^2$. This is their eq. (19),
\begin{equation}\label{eq:pl-quartic-factored}
\begin{aligned}
&[\, \omega^4 + \omega^3 |E-I| - \omega^2 (k^2 + EI + 1) - \omega k^2 |E-I| + k^2 EI \,] \\
\times &[\, \omega^4 - \omega^3 |E-I| - \omega^2 (k^2 + EI + 1) + \omega k^2 |E-I| + k^2 EI \,] = 0.
\end{aligned}
\end{equation}
Since applying the transformation $\omega \rightarrow -\omega$ to either factor results in the other one, this expression mixes forward and backward propagating wave pairs. It was already argued in Ref. \onlinecite{Keppens2019_coldei} that this factorisation does not offer any advantage over (\ref{eq:branch-pl4}). Nevertheless, this factorisation is omnipresent in the literature. This is shown in Sec. \ref{sec:kinetic}.

\subsection{Perpendicular propagation}
Substituting $\lambda = 1$ to describe perpendicular propagation makes the dispersion relation factorise. Due to its length, the perpendicular dispersion relation has been moved to App. \ref{app:disprel} and is given by eq. (\ref{eq:disp-perp}). From MHD it is known that the S (slow) and A (Alfv\'en) modes do not propagate perpendicular to the magnetic field. Hence, we see that these two modes factor out as $\omega^4$. The non-zero modes are described by a linear branch and a cubic branch in $\omega^2$. Note that the cubic branch contains both magnetic characteristics ($E$, $I$) and acoustic characteristics ($v$, $w$) unlike the branches at parallel propagation.

The linear branch is clearly an electromagnetic wave with a cutoff of $1$. Since the highest cutoff ($\omega_\mathrm{u} > 1$) always corresponds to the X mode, this must be the O mode. This is also in accordance with the definition of the ordinary mode in the literature as the mode that does not depend on the magnetic field \cite{GurnettBhattacharjee2005}. The remaining three modes, F, M, and X, are thus described by the cubic branch.

\section{Dispersion diagrams}\label{sec:dispdiagrams}
Once a choice is made for the parameters $E$, $\mu$, $v$, and $w$, the six branches $\omega(k)$ can be computed numerically at any angle ($\lambda^2\in [ 0,1 ]$). From MHD it is known that at long wavelengths (small $k$) the slow, Alfv\'en, and fast frequencies are always ordered as $\omega_\mathrm{S} \leq \omega_\mathrm{A} \leq \omega_\mathrm{F}$. Additionally, the frequencies of the M, O, and X modes lie above these with their cutoffs at $1$ and $\omega_\mathrm{u,l}$. It was shown that for pair plasmas and cold ion-electron plasmas a complete ordering of modes $\omega_\mathrm{S} \leq \omega_\mathrm{A} \leq \omega_\mathrm{F} \leq \omega_\mathrm{M} \leq \omega_\mathrm{O} \leq \omega_\mathrm{X}$ is satisfied at all angles except for parallel and perpendicular propagation \cite{Keppens2019_coldpair, Keppens2019_coldei, Keppens2019_warmpair}. At these extreme angles, the ordering is violated due to the introduction of crossings inbetween the long and short wavelength limits. In this section the crossings are discussed and visualised for parallel and perpendicular propagation for the warm ion-electron case. Afterwards, it is shown that at intermediate angles no crossings occur such that in the warm ion-electron case the modes are ordered as well.

\subsection{Parallel propagation}\label{sec:dispdiagram-pl}
As discussed in Sec. \ref{sec:disprel-pl}, at parallel propagation the dispersion relation splits into two factors, a quadratic and a quartic branch. It was already mentioned that the quadratic branch is the same as in the unmagnetised case whilst the quartic branch appears in the cold case as well. This means that the four modes described by the quartic branch are not influenced by the thermal speeds. Hence, the cold and warm dispersion diagrams will look similar.

\begin{figure*}[!htb]
\centering
\includegraphics[width=\textwidth]{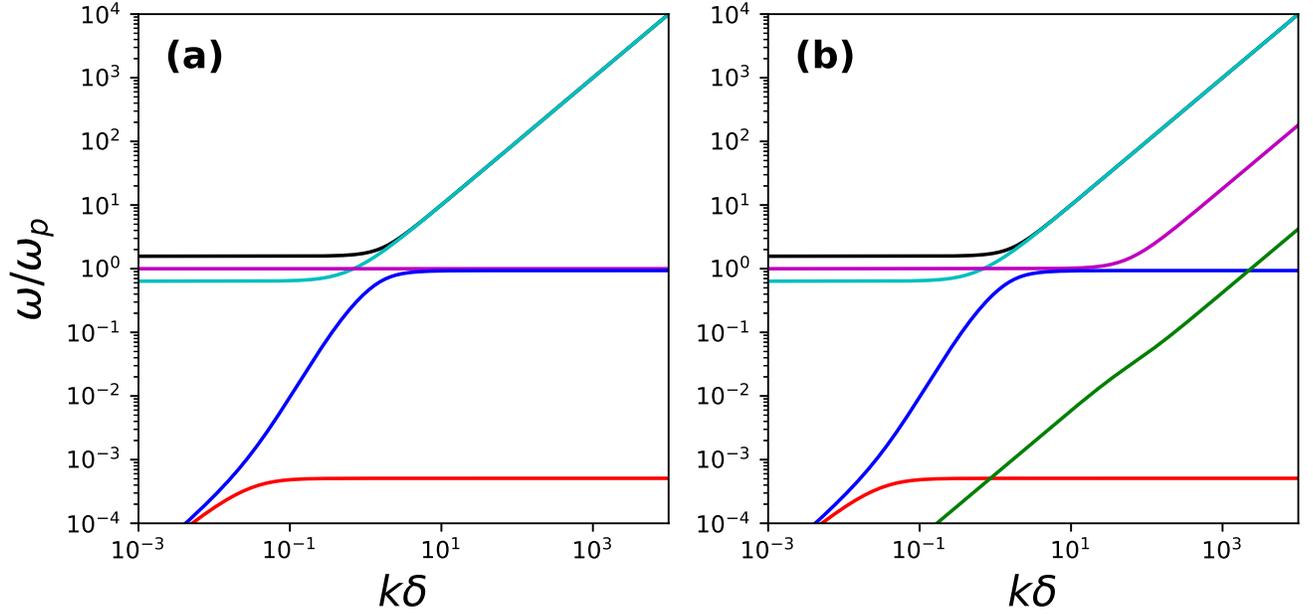}
\caption{Parallel dispersion diagram using coronal loop parameters \cite{GoedbloedKeppensPoedts2019} ($E \simeq 0.935$, $\mu \simeq 1/1836$) for (a) a cold ion-electron plasma, and (b) a warm ion-electron plasma ($v \simeq 0.018$, $w \simeq 0.0004$). Note that the A (red), F (blue), O (cyan), and X (black) modes are identical in the cold and warm case. The M (purple) mode's short wavelength (large $k$) behaviour is altered and the S (green) mode only appears in the warm case.}
\label{fig:warmcold-comparison}
\end{figure*}

In the cold case, the remaining mode is a constant mode \cite{Keppens2019_coldei} which feels the influence of the electron sound speed in the warm case. Consequently, its high-frequency behaviour is altered dramatically. This is the textbook Langmuir wave \cite{GurnettBhattacharjee2005} or M mode. Additionally, the slow mode appears. To visualise these differences, Fig. \ref{fig:warmcold-comparison} offers a side-by-side view of the cold and warm parallel dispersion diagrams. (In Ref. \onlinecite{Keppens2019_coldei} they called the constant mode the O mode. Due to the differing convention it is called the M mode here instead. Fig. \ref{fig:warmcold-comparison} follows our convention colour-wise.)

A brief glance at the right frame of Fig. \ref{fig:warmcold-comparison} suffices to notice several modes crossing. In fact, the case shown there is only one possible regime in which crossings appear ($E < 1$, $c_\mathrm{s} < c_\mathrm{a}$). Which modes cross is determined by the strength of the magnetic field, the sound speed, and the ratio of masses over charges. Eight regimes can be identified. They are characterised by the value of $E$ ($E < 1$, $1 < E < E_{\mathrm{cr}}$, $E_{\mathrm{cr}} < 1/\mu$, or $E > 1/\mu $) and whether $c_\mathrm{s} < c_\mathrm{a}$ or $c_\mathrm{s} > c_\mathrm{a}$. However, if $E$ exceeds $E_{\mathrm{cr}}$ for a realistic value $\mu \lesssim 1/1836$, the relation $c_\mathrm{s} > c_\mathrm{a}$ is no longer physically feasible because the relativistic sound speed is bounded by $c_\mathrm{s}^2 < \gamma-1$ ($\gamma = 5/3$, or $\gamma = 4/3$ in a relativistic regime) \cite{GoedbloedKeppensPoedts2019}. Hence, we limit ourselves to the remaining six cases here. A summary of all crossings in all regimes is given in Table \ref{table:pl-crossings}. 

\begin{table}[!htb]
\centering
\caption{Overview of crossings at parallel propagation in different regimes as determined using both analytical and numerical methods. Each crossing is indicated by the two letters corresponding to the crossing modes.}
\label{table:pl-crossings}
\begin{tabular}{ r | l l } 
& $c_\mathrm{s} < c_\mathrm{a}$ & $c_\mathrm{s} > c_\mathrm{a}$ \\
\hline
$E < 1$ & MO, SA, SF & MO, AF ($2 \times$) \\[1ex]
$1 < E < E_{\mathrm{cr}}$ & MO, SA, SF, FM ($2 \times$) & MO, AF ($2 \times$), AM ($2 \times$) \\[1ex]
$E_{\mathrm{cr}} < E < 1/\mu$ & SA, SF, FM ($2 \times$) & unphysical \\[1ex]
$E > 1/\mu$ & SA, SF, FM ($2 \times$), AM ($2 \times$) & unphysical
\end{tabular}
\end{table}

To discuss analytical expressions of these crossings, a closer look at the quadratic branch is required. Due to its quadratic nature, the expressions for $\omega^2$ as a function of $k^2$ can be written explicitly, namely 
\begin{equation}\label{eq:freq-q2l}
\begin{aligned}
\omega^2 = &\frac{1}{2} + \frac{1}{2}k^2 (v^2+w^2) \\ &\pm \frac{1}{2} \sqrt{k^4(v^2-w^2)^2 + 2k^2(v^2+w^2-2c_\mathrm{s}^2) + 1}.
\end{aligned}
\end{equation}
In eq. (\ref{eq:freq-q2l}), the solution with the positive sign is the M mode as can be seen by taking the short wavelength limit, $\omega^2 = k^2v^2$. This leaves the negative sign to describe the S or F mode because the long wavelength limit is $\omega^2 = k^2 c_\mathrm{s}^2$ (see Sec. \ref{sec:glf-limit}). From eq. (\ref{eq:freq-q2l}) it is also immediately clear that these two modes can never cross and could only coincide if the square root became zero. Observing that the expression underneath the square root is a quadratic polynomial in $k^2$, we can compute the discriminant $\Delta = (c_\mathrm{s}^2-v^2)(c_\mathrm{s}^2-w^2)$. Since $\min\{v^2,w^2\} \leq c_\mathrm{s}^2 \leq \max\{v^2,w^2\}$, this is always negative and thus there are no real crossings of the modes in the quadratic branch. The quartic branch was discussed in Ref. \onlinecite{Keppens2019_coldei} and no crossings appear within this branch either.

\begin{figure*}[!htb]
\centering
\includegraphics[width=\textwidth]{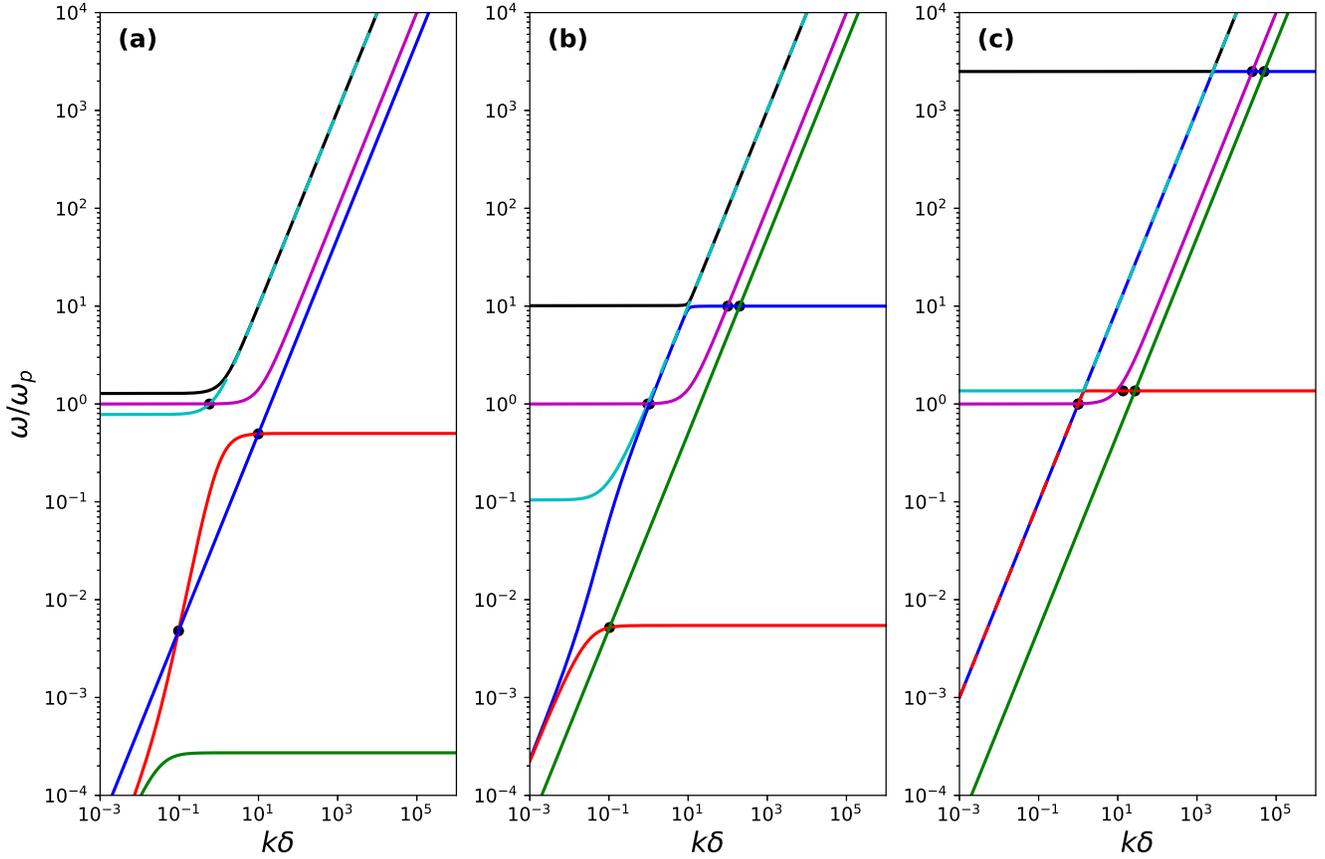}
\caption{Dispersion diagram of a proton-electron plasma at parallel propagation with increasing magnetic field strength from left to right: (a) $E = 0.5$, (b) $E = 10$, and (c) $E = 2500$. All three cases satisfy $\mu \simeq 1/1836$, $v = 0.1$ and $w = 0.05$. For increasing magnetic field strength new crossings appear of the M mode with the F and A modes. Above $E_{\mathrm{cr}}$, in (c), the M and O modes no longer cross. Analytical approximations of the crossings are marked by a black dot. Dashed lines are used when modes are too close to discern otherwise.}
\label{fig:pl_Evar}
\end{figure*}

The remaining question is when and where the quadratic branch crosses the quartic branch. In Table \ref{table:pl-crossings} we already summarised when the modes cross. Unfortunately, a first attempt to solve the quadratic and quartic branch as a system in the variables $\omega^2$ and $k^2$ did not yield simple, closed-form, analytical results. However, with the use of a Taylor expansion in eq. (\ref{eq:freq-q2l}) it is possible to retrieve approximate analytical solutions for the locations of the crossings. In Fig. \ref{fig:pl_Evar} three regimes are shown as a reference for the analytical approximations that follow. All analytical approximations are marked in this figure with a black dot.

First off, consider the crossing of the M and O modes. These modes cross at parallel propagation if $E < E_{\mathrm{cr}}$ and can be seen in panels a and b of Fig. \ref{fig:pl_Evar} in purple (M) and cyan (O). Analytically, this is clear from observing that the short wavelength limit of the M mode $\omega^2 = k^2v^2$ is smaller than the corresponding limit of the O mode $\omega^2 = k^2$ whilst the cutoff of the M mode $\omega^2 = 1$ is larger than the cutoff of the O mode $\omega^2 = \omega_\mathrm{l}^2$ if $E < E_{\mathrm{cr}}$. Assuming the crossing occurs at relatively small $k$ as suggested by Fig. \ref{fig:pl_Evar}, $\omega_\mathrm{M}^2 = 1$ to zeroth order. Unfortunately, retaining any term of first or higher order in $k^2$ makes the substitution in the quartic branch a lot more involved. The equation would become fourth order in $k^2$ which is analytically solvable, but becomes extremely lengthy. To keep it simple, substituting $\omega^2 = 1$ in the quartic branch and solving for $k^2$ gives for the parallel crossing of the M and O mode
\begin{equation}\label{eq:plcross-approx-cst}
k^2_{\mathrm{MO},\parallel} \simeq \frac{EI - (E-I)}{EI-1 - (E-I)}.
\end{equation}
Since $\omega^2 = 1$ is an exact solution in the cold ion-electron case, this expression also appears there as an exact crossing \cite{Keppens2019_coldei}. Comparing this approximation of the crossing $(\omega, k) = (1, k_{\mathrm{MO},\parallel})$ to numerical results shows that this approximation is usually quite good.

For the lower solution ($-$) of eq. (\ref{eq:freq-q2l}), using a first order Taylor approximation of the square root and discarding the second order term in $k^2$, the mode's behaviour becomes $\omega^2 = k^2c_\mathrm{s}^2$. Substituting this simple expression into the quartic branch and solving for $k^2$ yields
\begin{equation}\label{eq:plcross-approx-lin}
\begin{aligned}
k^2 \simeq &\left[ 2c_\mathrm{s}^2(1-c_\mathrm{s}^2) \right]^{-1} \Big\{ (E^2+I^2)(1-c_\mathrm{s}^2)-2c_\mathrm{s}^2 \\
&\pm |E-I| \sqrt{(1-c_\mathrm{s}^2)\left[ (E+I)^2 (1-c_\mathrm{s}^2) - 4c_\mathrm{s}^2 \right]} \Big\}.
\end{aligned}
\end{equation}
This expression can give either $0$, $1$, or $2$ real, positive solutions. The solutions agree fairly well with numerical solutions. To discuss which crossings this expression describes, we have to differentiate between the two regimes $c_\mathrm{s} < c_\mathrm{a}$ and $c_\mathrm{s} > c_\mathrm{a}$.

The first regime, $c_\mathrm{s} < c_\mathrm{a}$, is shown in panels b and c of Fig. \ref{fig:pl_Evar}. Here, the lower mode of the quadratic branch is the S mode which crosses both the A and F mode once. In this case the smaller solution of eq. (\ref{eq:plcross-approx-lin}) gives the SA crossing whereas the larger solution gives the SF crossing. In the other regime, $c_\mathrm{s} > c_\mathrm{a}$, the lower quadratic branch solution is the F mode which starts above both the S and A mode. This case is shown in panel a of Fig. \ref{fig:pl_Evar}. Since $\omega_\mathrm{S} < \omega_\mathrm{A} < \omega_\mathrm{F}$ holds in both the long and short wavelength limits and there are at most two crossings, it follows that the F mode never crosses the S mode, but it can cross the A mode twice with each solution of eq. (\ref{eq:plcross-approx-lin}) describing an AF crossing.

It should be pointed out that the upper solution in eq. (\ref{eq:plcross-approx-lin}) can become rather large. In this case using a small $k$ approximation is questionable at best. However, note that the short wavelength limit of this mode is $\omega^2 = k^2w^2$ and is linear like the long wavelength limit, but with a different coefficient. This means that if the crossing appears at high $k$ we can simply replace $c_\mathrm{s}$ with $w$ in eq. (\ref{eq:plcross-approx-lin}).

Finally, two more crossings appear for values of $E$ larger than $E > 1$ and another two for $E$ larger than $E > 1/\mu$ (with $c_\mathrm{s} < c_\mathrm{a}$). When $E$ becomes larger than $1$, the largest resonance $\omega^2 = E^2$ exceeds the M mode cutoff $\omega^2 = 1$. Then the M mode will cross the mode approaching this resonance twice. If $c_\mathrm{s} < c_\mathrm{a}$, this means the M and F modes cross. If $c_\mathrm{s} > c_\mathrm{a}$ on the other hand, the M and A modes cross instead. As an aside, note that in the latter case this means that the A mode crosses four times with other modes, in the order F, M, M and F for $k$ ranging from $0$ to $+\infty$. Additionally, the F and M modes never cross if $c_\mathrm{s} > c_\mathrm{a}$.

The way we can approximate these crossings is by replacing the frequency in the quartic branch with the long and short wavelength limits of the M mode, i.e. $\omega_\mathrm{M}^2 \simeq 1$ for the first crossing and $\omega_\mathrm{M}^2 \simeq k^2v^2$ for the second crossing. For the first crossing this gives
\begin{equation}\label{eq:pl-cross-largeE}
k^2 \simeq \frac{EI + (E-I)}{EI-1 + (E-I)}.
\end{equation}
Once again, this is an exact solution in the cold ion-electron case \cite{Keppens2019_coldei}. In our test cases this yields an acceptable approximation. The approximation for the second crossing is given by replacing $c_\mathrm{s}$ with $v$ in eq. (\ref{eq:plcross-approx-lin}). However, for  values of $E$ relatively close to (but larger than) $E=1$ ($1 < E \lesssim 10$ in our test cases where $\mu = 1/1836$), the second crossing falls into the transitional regime between the long and short wavelength limits of the M mode which results in a rather suboptimal approximation. For large values of $E$ ($E \gtrsim 10$) the approximation is exceptionally good though.

If $E$ exceeds $E > 1/\mu$, the lower resonance $\omega^2 = I^2 = \mu^2 E^2$ also becomes greater than the M mode cutoff $\omega_\mathrm{M}^2 = 1$. Since $c_\mathrm{s} < c_\mathrm{a}$, the A mode crosses the M mode twice. As can be seen in the right panel of Fig. \ref{fig:pl_Evar}, for strong magnetic fields the F and A mode almost coincide such that the same approximation (\ref{eq:pl-cross-largeE}) can be used for both the AM and FM crossing. For the second crossing, the short wavelength limit $\omega_\mathrm{M}^2 = k^2v^2$ can be used to find the lower solution eq. (\ref{eq:plcross-approx-lin}) with $c_\mathrm{s}$ replaced by $v$. Once again, this is not a great approximation because the M mode might not be entirely in the short wavelength regime.

All approximations offered in this section were tested against numerical results of the crossings using test cases for each regime. These test cases are differentiated by their parameters and can be found in App. \ref{app:testcases}. For each case the numerical crossings and their analytical approximations are listed.

\subsection{Perpendicular propagation}\label{sec:dispdiagram-pp}
As shown before, the perpendicular case splits into two branches. The linear branch $\omega^2 = 1+k^2$ is the O mode and the remaining cubic branch describes the F, M, and X modes.

\begin{figure*}[!htb]
\centering
\includegraphics[width=\textwidth]{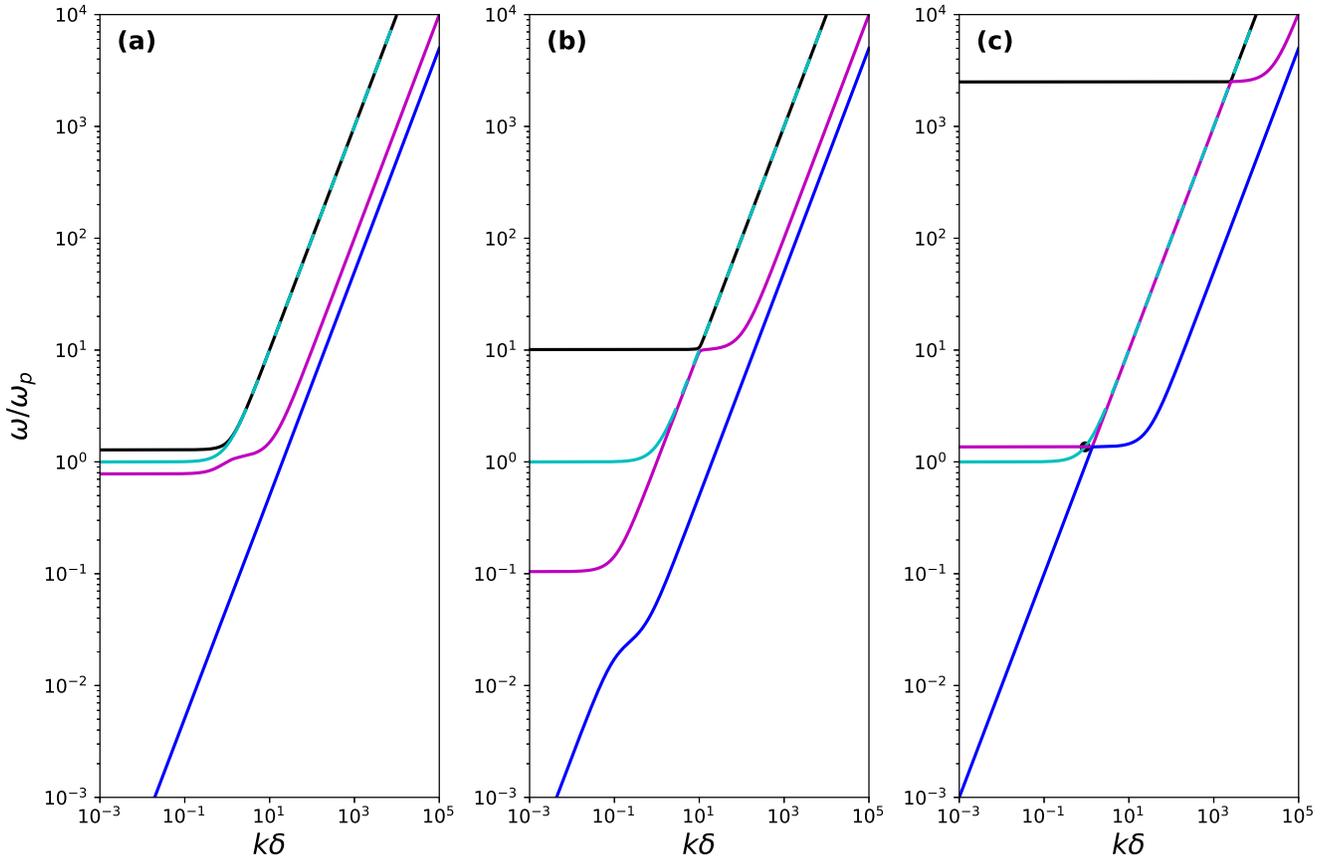}
\caption{Dispersion diagram of a proton-electron plasma at perpendicular propagation with increasing magnetic field strength from left to right: (a) $E = 0.5$, (b) $E = 10$, and (c) $E = 2500$. All three cases satisfy $\mu \simeq 1/1836$, $v = 0.1$ and $w = 0.05$. Above $E_{\mathrm{cr}}$, in (c), a new crossing appears between the M and O modes. The analytical expression of the crossing is marked by a black dot. Dashed lines are used when modes are too close to discern.}
\label{fig:pp_Evar}
\end{figure*}

For the perpendicular case the issue of crossings is less involved than in the parallel case mainly because there are less propagating modes. There are no crossings or avoided crossings within the cubic branch, so the only remaining question is if the linear branch crosses any of the modes in the cubic branch. Substituting $\omega^2 = 1+k^2$ into the cubic branch and solving for $k^2$, the solution can be written as
\begin{equation}\label{eq:crossing-perp}
k_{\mathrm{MO},\perp}^2 = \frac{\mu^2(1+\mu) \left( E^2 - E_{\mathrm{cr}}^2 \right)}{(1-w^2) + \mu(1-v^2)}.
\end{equation}
The subscript MO indicates that this is the wavenumber of the crossing of the M and O modes. Now it should be noted that the denominator is always positive whilst the numerator goes from a negative to a positive value for increasing $E$ at $E = E_{\mathrm{cr}}$. Therefore, a real crossing between the M and O modes appears if $E > E_{\mathrm{cr}}$ at the indicated wavenumber value. The corresponding frequency is simply $\omega_{\mathrm{MO},\perp}^2 = 1+k_{\mathrm{MO},\perp}^2$. For $E = E_{\mathrm{cr}}$ the ``crossing" is the cutoff point $(\omega^2 = 1, k^2=0)$. The perpendicular dispersion diagram is shown in Fig. \ref{fig:pp_Evar}.

As a final check we can compare to the cold pair plasma case. In Ref. \onlinecite{Keppens2019_coldpair} they reported an MO crossing at perpendicular propagation at $(\omega, k) = (\sqrt{1+E^2}, E)$. As noted before, for a cold pair plasma the ratio of masses over charges is $\mu = 1$ and the critical electron cyclotron frequency is $E_{\mathrm{cr}} = 0$ as well as the sound speeds $v = 0$ and $w = 0$. These simplifications reduce eq. (\ref{eq:crossing-perp}) to $k = E$ such that we recover the cold pair plasma result.

\subsection{Oblique propagation}\label{sec:dispdiagram-obl}
For cold and warm pair plasmas \cite{Keppens2019_coldpair, Keppens2019_warmpair} as well as cold ion-electron plasmas \cite{Keppens2019_coldei} crossings are replaced by avoided crossings as soon as there is any deviation from parallel or perpendicular propagation. Consequently, this means that the connectivity between long and short wavelength limits is angle dependent. As it turns out, similar changes occur in a warm ion-electron plasma. At oblique angles, modes almost surely do not cross, and the ordering
\begin{equation*}
\omega_\mathrm{S} \leq \omega_\mathrm{A} \leq \omega_\mathrm{F} \leq \omega_\mathrm{M} \leq \omega_\mathrm{O} \leq \omega_\mathrm{X}
\end{equation*}
is obeyed at all wavelengths. The fact that this ordering of the eigenfrequencies persists through all oblique orientations makes this a more robust basis for categorising all waves in ion-electron plasmas. It also forms a natural extension of the five waves found in the cold limit and remains true for all realistic values of $\mu$ (up to the pair plasma limit where $\mu=1$). We therefore argue that it should replace previous labelling schemes, which actually mixed wave labels due to insisting to connect parallel to perpendicular wave properties, and/or confused forward and backward wave pairs as resulting from artificial factorisations as given by eq. (\ref{eq:pl-quartic-factored}).

\begin{figure*}[!htb]
\centering
\includegraphics[width=\textwidth]{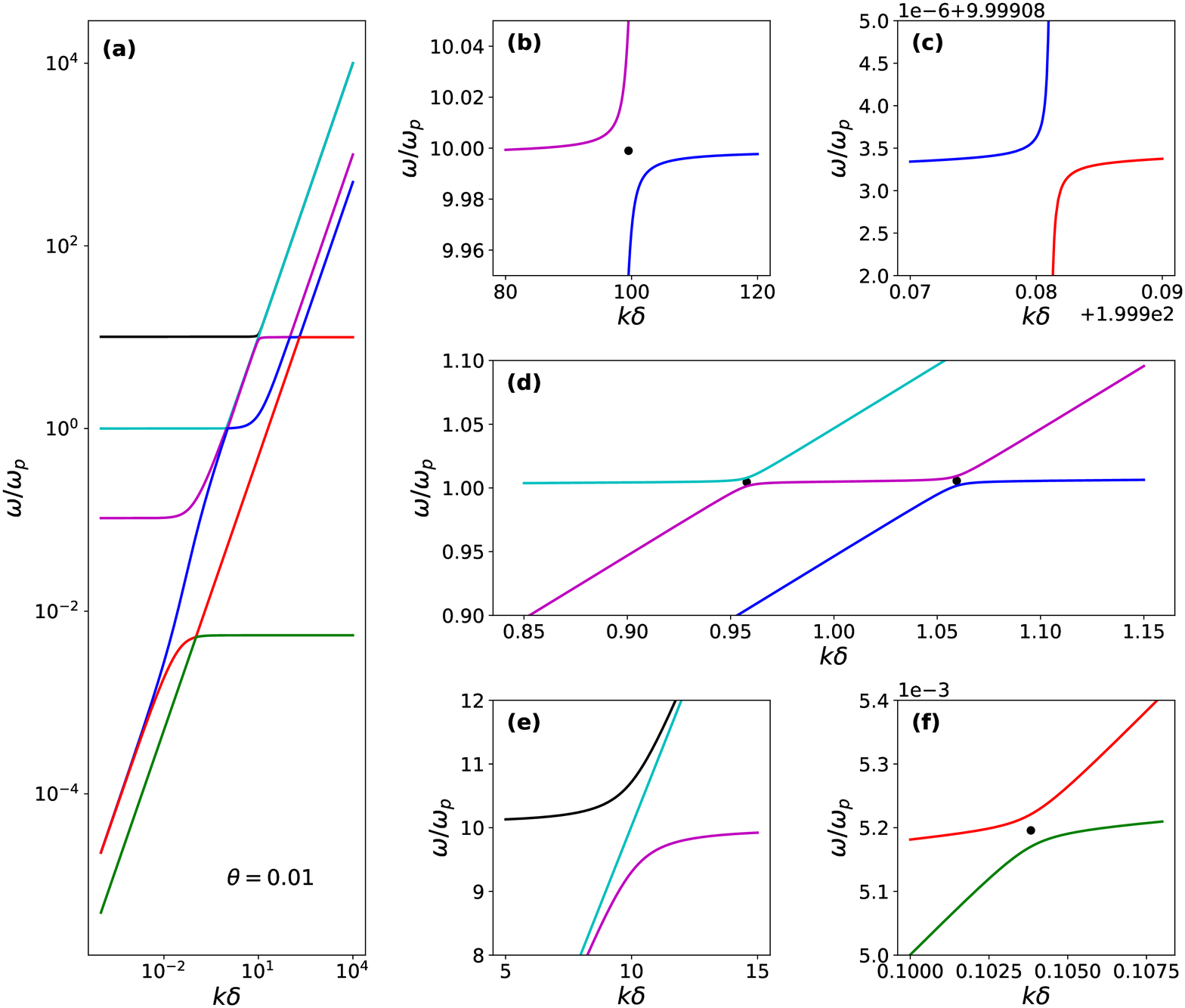}
\caption{(a) A proton-electron dispersion diagram for nearly-parallel propagation ($\theta = 0.01$) using the parameters of Fig. \ref{fig:pl_Evar}(b). The close-ups show the (b) FM region, (c) AF region, (d) FMO region, (e) MOX region, and (f) SA region. All crossings are replaced by avoided crossings. The numerical crossings at parallel propagation are indicated by black dots. For the AF crossing, close-up in (c), the numerical result at parallel propagation lies outside of the frame, $(\omega, k) = (9.99974909, 199.83187597)$. The MOX region, close-up in (e), has no crossing at parallel propagation.}
\label{fig:avoidedcrossings}
\end{figure*}

As soon as we move away from exactly parallel propagation, \emph{all} crossings that were present at parallel propagation seem to turn into avoided crossings. The avoided crossings are shown in Fig. \ref{fig:avoidedcrossings} for a small angle. This change from crossing to avoided crossing with the smallest deviation from parallel propagation gives the parallel case a special status. This is actually worrisome for classical textbook treatments where purely parallel (or purely perpendicular) orientations serve to identify and categorise mode properties. The location of these avoided crossings for small angles can be pinpointed by computing the crossings at parallel propagation numerically or they can be approximated using the crossing expressions in Sec. \ref{sec:dispdiagram-pl}.

However, for the AF crossing the location of the avoided crossing away from parallel propagation does not match the numerical crossing at parallel propagation. In fact, from numerical results it seems that the location of this crossing is extremely angle sensitive. When varying the angle $\theta$ (and thus $\lambda$), the point of closest approach between the two modes, i.e. the avoided crossing, varies rapidly as well. Even for small angles close to parallel propagation the distance between the crossing at parallel propagation and the avoided crossing is significantly larger than the closest distance between the two modes in the avoided crossing. (To test this we could only go up to $\theta = 0.001$ because smaller values were no longer numerically well-resolved.) As can be seen in panel c of Fig. \ref{fig:avoidedcrossings}, for an angle $\theta = 0.01$ the numerical crossing at parallel propagation is already well outside of the avoided crossing's frame. Note the scale on the axes though, since the A and F modes approach each other closely.

For the perpendicular MO crossing in the $E > E_{\mathrm{cr}}$ regime, the situation is a little different. As soon as the angle deviates from $\pi/2$, the S and A mode are reintroduced. Due to their appearance the MO crossing does not simply become an avoided crossing, but several modes approach each other closely without crossing. The result looks a lot like the parallel case in panel c of Fig. \ref{fig:pl_Evar} with all crossings replaced by avoided crossings. All six modes are present and the avoided crossings related to crossings at parallel propagation persist through all angles up to near-perpendicular propagation. Recolouring the branches appropriately reveals that the avoided crossings are SA, FM twice and AF three times. Since these avoided crossings actually originated from crossings at parallel propagation, comparing them to Table \ref{table:pl-crossings} reveals that the (parallel) SF crossing and both AM crossings became AF avoided crossings. This is simply due to the fact that modes can no longer cross. Thus, the labelling of avoided crossings at higher wavenumbers can be affected by avoided crossings at smaller wavenumbers. As an example, consider the parallel SF crossing, visible in Fig. \ref{fig:pl_Evar}b and its corresponding nearly-parallel case in Fig. \ref{fig:avoidedcrossings}a,c. The S mode approaches the F mode because it already crossed the A mode at a smaller wavenumber. However, when that SA crossing becomes an avoided crossing, it is now the A mode (instead of the S mode) that approaches the F mode, so the SF crossing becomes an AF avoided crossing. Consequently, due to the S, A, F, M, O, X frequency ordering at oblique angles, avoided crossing labels are always made up of two consecutive modes in this order, e.g. SA, AF, and FM avoided crossings, whilst this is not necessarily the case for the true crossings at parallel propagation, e.g. SF and AM crossings, where the S and A mode already crossed the A and F modes respectively at smaller wavenumbers.

\subsection{Critical electron cyclotron frequency}\label{sec:Ecr}
\begin{figure*}[!htb]
\centering
\includegraphics[width=\textwidth]{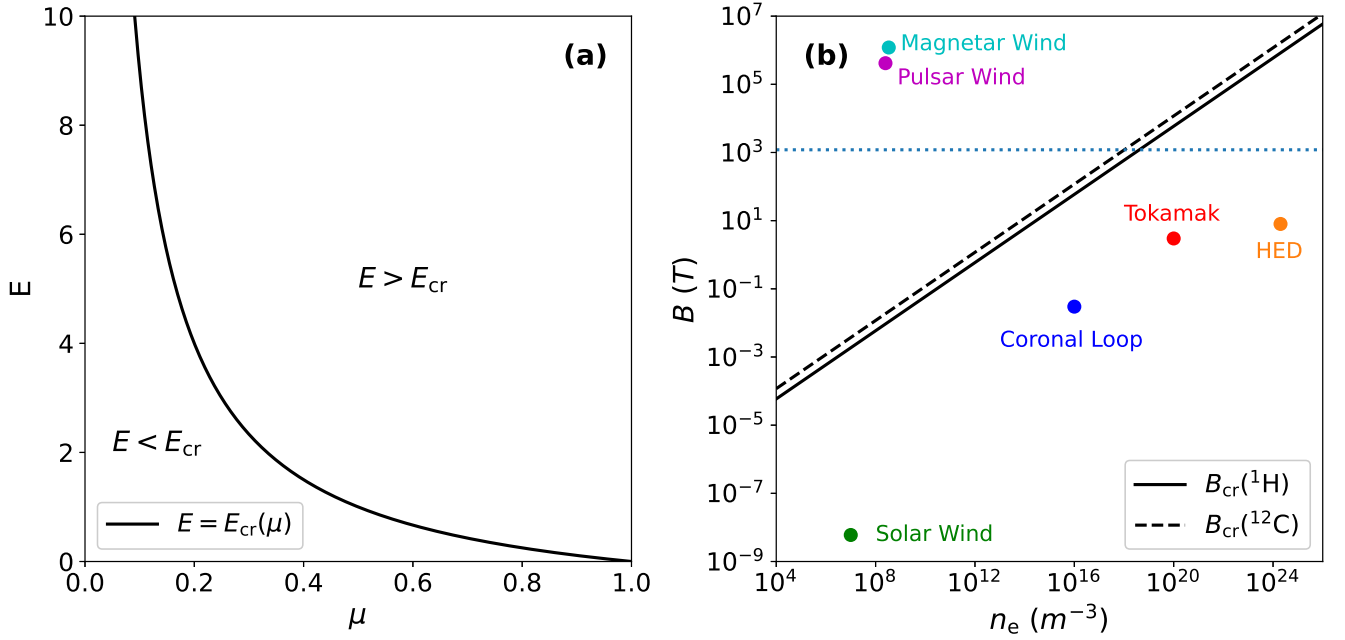}
\caption{(a) The critical value $E_{\mathrm{cr}}$ as a function of the ratio of masses over charges $\mu$. For a pair plasma ($\mu = 1$) the critical value goes to zero whilst for an ion-electron plasma ($\mu \ll 1$) it becomes very large. (b) The magnetic field strength for which the critical value $E_{\mathrm{cr}}$ would be attained for a proton-electron plasma (solid line) and a carbon-electron plasma (dashed line) as functions of the (electron) density. A selection of physical environments are represented by a dot. \cite{GoedbloedKeppensPoedts2019, Petri2019, GoldreichJulian1969, Fiksel2014, Fox2017} The dashed blue line indicates the record artificial magnetic field strength, $B = 1200\ \mathrm{T}$. \cite{Nakamura2018}}
\label{fig:EcrBcr}
\end{figure*}

Clearly, the critical electron cyclotron frequency $E_{\mathrm{cr}}$ plays an important role in the behaviour of the different wave types. When $E$ crosses the threshold imposed by the critical value, crossings between the M and O modes appear or disappear at parallel and perpendicular propagation. Consequently, at parallel and perpendicular propagation the ordering of the M and O modes switches at $k^2 = 0$ across the critical value. On the other hand, for oblique propagation the ordering is fixed and their cutoff expression are interchanged. This distinction based on the strength of the magnetic field was absent in the case of a warm pair plasma \cite{Keppens2019_warmpair} because there the critical value is $E_{\mathrm{cr}} = 0$. Since the critical value only depends on $\mu$, the function can be shown in a 2D plot as is done in Fig. \ref{fig:EcrBcr}a.

Whilst it is clear that it is zero for a pair plasma, it rapidly increases for smaller values of $\mu$. Therefore, a pair plasma is always in the upper regime, but there are two possible regimes for any ion-electron plasma ($\mu \ll 1$). However, as Fig. \ref{fig:EcrBcr}b shows, the upper regime is rather extraordinary for an ion-electron plasma due to the necessity of immensely strong magnetic fields. In this figure the critical magnetic field strength $B_{\mathrm{cr}}$ corresponding to $E_{\mathrm{cr}}$ is plotted for a proton-electron plasma ($\mu \simeq 1/1836$) and a carbon-electron plasma as a function of the number density. A couple of physical cases have been added to the figure, both for the lower and the upper regime. Noting the logarithmic scale of the figure, it is clear that for many regular cases the magnetic field is several orders of magnitude below the critical value \cite{GoedbloedKeppensPoedts2019}. The strongest artificially generated magnetic fields are employed in high-energy density experiments and an HED carbon-electron plasma experiment \cite{Fiksel2014, Fox2017} is also shown for reference, although it is still well below the critical value. However, for more extreme cases such as pulsars and magnetars, the magnetic field exceeds the critical value for their proton-electron wind. Estimates of the magnetic field $B_w$ in Ref. \onlinecite{Petri2019} were used alongside Goldreich-Julian density estimates \cite{GoldreichJulian1969} for the pulsar (J1734-3333) and magnetar (Swift J1834) cases in Fig. \ref{fig:EcrBcr}b. Considering that the strongest artificial magnetic field had a strength of $1200\ \mathrm{T}$, \cite{Nakamura2018} applying such a field to a low-density plasma may create a suitable environment to study this upper regime experimentally.

\section{Phase and group speed}
Naturally, each wave has two speeds associated with it, namely the phase and group speed. Since the dispersion relation is a sixth order polynomial in $\omega^2$, it is not possible in general to write down an analytical expression for these speeds for a given mode as a function of the wavenumber $k$. In the case of parallel or perpendicular propagation, the sixth order dispersion relation reduces to factors of at most fourth order, which are analytically solvable for functions $\omega(k)$. However, the resulting phase and group speed expressions are too involved to show. Moreover, if the pair $(\omega,k)$ is determined numerically, these values can be used to compute the phase and group speed at any angle. Hence, with the use of numerical methods it is more convenient to show the results in phase and group speed diagrams which will collect the information of plane wave propagation speeds (phase diagram) or energy flow (group diagram) for all angles at once at a given wavenumber $k$ (hence given wavelength). As the dispersion relations showed earlier that all 6 waves are dispersive, i.e. all $\omega(k)$ branches deviate from a mere proportionality $\omega(k)\propto k$, and this essentially at all angles, the phase and group speeds that collect anisotropic wave propagation characteristics will differ for varying wavelengths. Thereby, the long and short wavelength limits can be computed analytically.

\subsection{Phase speed}\label{sec:phase-speed}
The (dimensionless) phase speed defined as $v_{\mathrm{ph}} = \omega / k$ can only be written as an explicit function of $k$ for each mode in the case of parallel or perpendicular propagation by solving the quadratic and quartic branch for functions $\omega(k)$. However, these expressions do not offer any insight and have been omitted here. In the long and short wavelength limit though, the expressions reduce significantly and can be found for all orientations between the wavevector and the magnetic field. Hence, these limit cases have been summarised in Table \ref{table:phasespeed}. Here, the unit vector $\hat{\mathbf{n}}$ is defined as $\hat{\mathbf{n}} = \mathbf{k}/k$. These expressions are in accord with the cutoff, resonance, MHD, and short wavelength limits discussed in Sec. \ref{sec:limitbehaviour}.

\begin{figure*}[!htb]
\centering
\includegraphics[width=\textwidth]{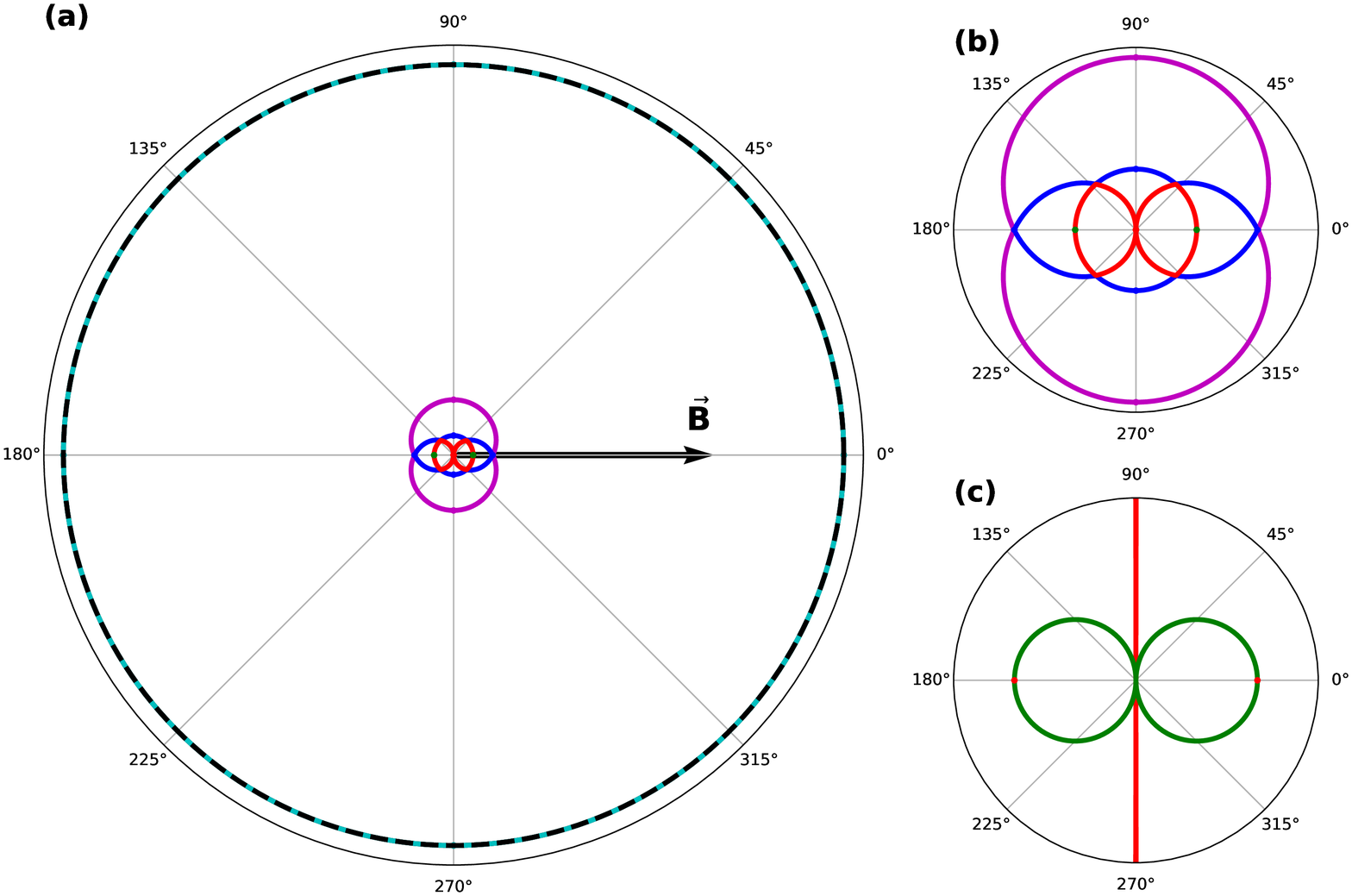}
\caption{(a) Phase speed diagram for $k = 100$ of a proton-electron plasma with $E=10$, $\mu = 1/1836$, $v=0.1$, and $w=0.05$. The dashed black line indicates the light circle. (b-c) Successively zoomed views of the central region of (a).}
\label{fig:phasespeed-large}
\end{figure*}

For all intermediate wavenumbers it is also possible to create complete phase speed diagrams. These diagrams show the phase speed in a polar plot, where the angle corresponds to the angle between the magnetic field and the wavevector. Fig. \ref{fig:phasespeed-large} is an example of such a diagram. In these diagrams the background magnetic field points to the right. The evolution of these diagrams whilst varying $k$ offers an interesting look at the structure of the various modes. Animations of the phase speed diagram evolving with the wavenumber are available as a supplement (see Supplementary materials).

\begin{table}[!htb]
\centering
\caption{Phase speeds of all modes in the short and long wavelength limit (large and small $k$ respectively) assuming $E < E_{\mathrm{cr}}$. All labels refer to oblique angles ($0 < \lambda < 1$).}
\label{table:phasespeed}
\begin{tabular}{ l l } 
\hline
Long wavelengths ($k \rightarrow 0$) & Short wavelengths ($k \rightarrow \infty$) \\[1ex]
\hline\\
$\displaystyle \left( \frac{\omega}{k} \right)_\mathrm{X} \hat{\mathbf{n}} = \frac{\omega_\mathrm{u}}{k} \hat{\mathbf{n}}$ & $\displaystyle \left( \frac{\omega}{k} \right)_\mathrm{X} \hat{\mathbf{n}} = \hat{\mathbf{n}}$ \\[2ex]
$\displaystyle \left( \frac{\omega}{k} \right)_\mathrm{O} \hat{\mathbf{n}} = \frac{1}{k} \hat{\mathbf{n}}$ &$\displaystyle \left( \frac{\omega}{k} \right)_\mathrm{O} \hat{\mathbf{n}} = \hat{\mathbf{n}}$ \\[2ex]
$\displaystyle \left( \frac{\omega}{k} \right)_\mathrm{M} \hat{\mathbf{n}} = \frac{\omega_\mathrm{l}}{k} \hat{\mathbf{n}}$ & $\displaystyle \left( \frac{\omega}{k} \right)_\mathrm{M} \hat{\mathbf{n}} = v \hat{\mathbf{n}}$ \\[2ex]
$\displaystyle \left( \frac{\omega}{k} \right)_\mathrm{F} \hat{\mathbf{n}} = v_\mathrm{f} \hat{\mathbf{n}}$ & $\displaystyle \left( \frac{\omega}{k} \right)_\mathrm{F} \hat{\mathbf{n}} = w \hat{\mathbf{n}}$ \\[2ex]
$\displaystyle \left( \frac{\omega}{k} \right)_\mathrm{A} \hat{\mathbf{n}} = \lambda c_\mathrm{a} \hat{\mathbf{n}}$ & $\displaystyle \left( \frac{\omega}{k} \right)_\mathrm{A} \hat{\mathbf{n}} = \frac{\lambda E}{k} \hat{\mathbf{n}}$ \\[2ex]
$\displaystyle \left( \frac{\omega}{k} \right)_\mathrm{S} \hat{\mathbf{n}} = v_\mathrm{sl} \hat{\mathbf{n}}$ & $\displaystyle \left( \frac{\omega}{k} \right)_\mathrm{S} \hat{\mathbf{n}} = \frac{\lambda I}{k} \hat{\mathbf{n}}$ \\[3ex]
\hline
\end{tabular}
\end{table}

\subsection{Group speed}\label{sec:group-speed}
Much like the warm pair plasma case discussed in Ref. \onlinecite{Keppens2019_warmpair} it is possible to write down a general expression for the group speeds of all the waves as a function of the pair $(\omega, k)$. Since the dispersion relation does not factorise in general, the group speed expression must be achieved with the use of differentiation with respect to $\mathbf{k}$ on the sixth degree polynomial dispersion relation. Doing so, the group speed can be written as
\begin{equation}\label{eq:groupspeed}
\frac{\partial\omega}{\partial\mathbf{k}} = -(v_{\mathrm{ph}} P_\omega)^{-1} \left[ P_k \hat{\mathbf{n}} + \frac{\lambda P_\lambda}{k^2} \left( \hat{\mathbf{b}} - \lambda\hat{\mathbf{n}} \right) \right].
\end{equation}
In this formula $\hat{\mathbf{n}}$ and $\hat{\mathbf{b}}$ are unit vectors defined as $\hat{\mathbf{n}} = \mathbf{k}/k$ and $\hat{\mathbf{b}} = \mathbf{B}/B$, $v_{\mathrm{ph}}$ indicates the phase speed as before, and $P_\omega$, $P_k$, and $P_\lambda$ are polynomials defined as
\begin{align}
P_\omega &= \sum\limits_{\substack{1 \leq m \\ 0\leq n \\ 3 \leq m+n \leq 6}} m \alpha_{mn}\, \omega^{2(m-1)} k^{2n}, \label{eq:p_om} \\
P_k &= \sum\limits_{\substack{0\leq m \\ 1 \leq n \\ 3 \leq m+n \leq 6}} n \alpha_{mn}\, \omega^{2m} k^{2(n-1)}, \label{eq:p_k} \\
\text{and}\quad P_\lambda &= \sum\limits_{\substack{0 \leq m,n \\ 3 \leq m+n \leq 6}} \frac{\partial\alpha_{mn}}{\partial\lambda^2} \, \omega^{2m} k^{2n}. \label{eq:p_lam}
\end{align}
It should be pointed out that these polynomials are defined somewhat differently than in Ref. \onlinecite{Keppens2019_warmpair}, but otherwise yield identical expressions. The details of these polynomials have been moved to App. \ref{app:group_polynom}.

\begin{figure*}[!htb]
\centering
\includegraphics[width=\textwidth]{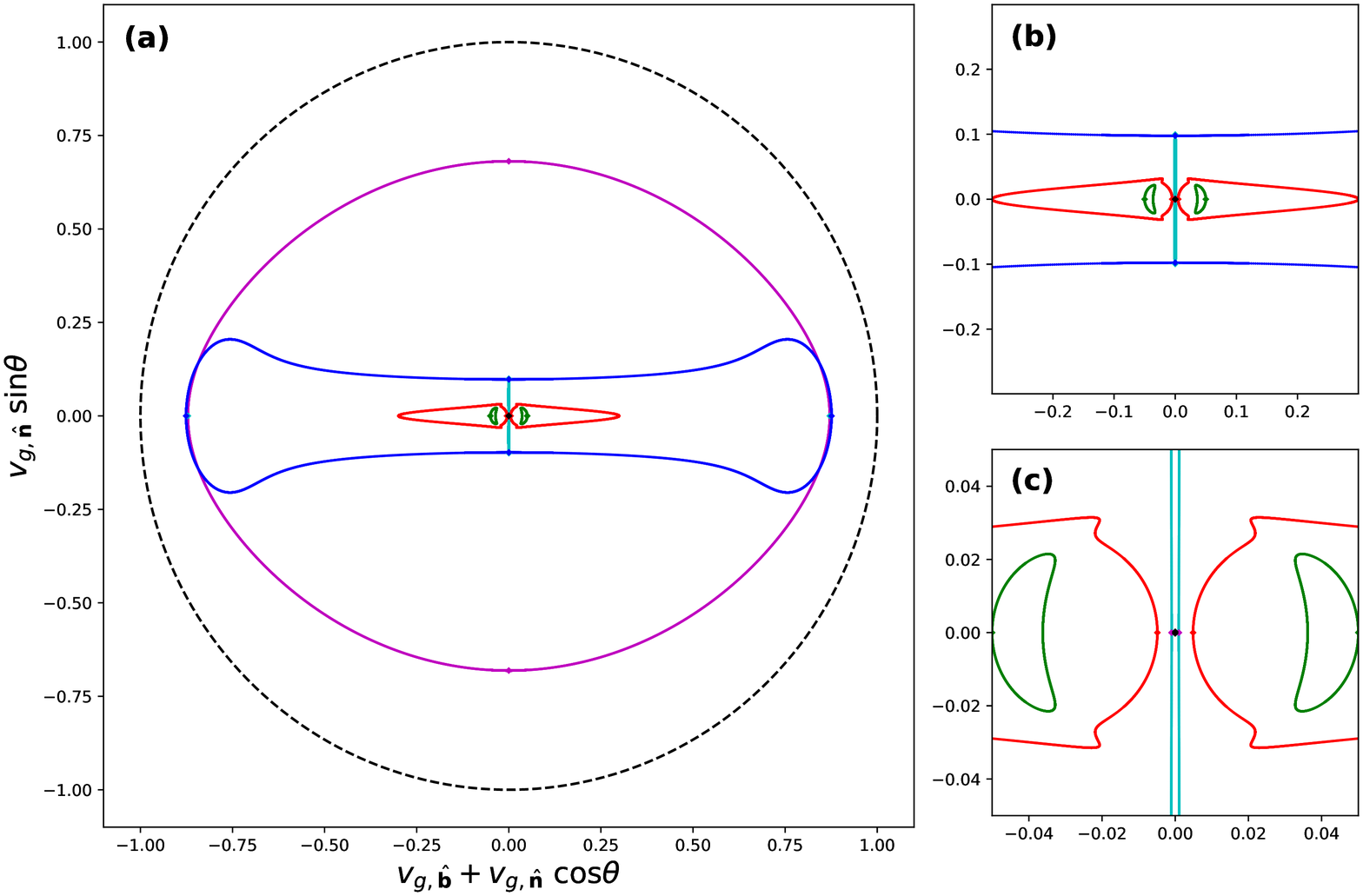}
\caption{(a) Group speed diagram for $k = 0.1$ of a proton-electron plasma with $E=10$, $\mu = 1/1836$, $v=0.1$, and $w=0.05$. The dashed black line indicates the light circle. (b-c) Successively zoomed views of the central region of (a).}
\label{fig:groupspeed}
\end{figure*}

The relation $\partial\lambda^2/\partial\mathbf{k} = 2\lambda (\hat{\mathbf{b}} - \lambda\hat{\mathbf{n}}) / k$ was invoked to reach expression (\ref{eq:groupspeed}). (Note that $\partial\lambda^2/\partial\mathbf{k} = 0$ at perpendicular ($\lambda = 0$) and parallel ($\lambda = 1$) propagation using $\hat{\mathbf{b}} = \lambda\hat{\mathbf{n}}$ for parallel propagation.) Of course, to obtain the group speed of any specific wave the pair $(\omega,k)$ solving the dispersion relation should be substituted into eq. (\ref{eq:groupspeed}). Analytically, this is only possible for parallel or perpendicular propagation, but group speeds can be computed numerically at any angle. In doing so, it is possible to draw group speed diagrams with $\hat{\mathbf{b}}$ pointing along the horizontal axis in the positive direction as is done in Fig. \ref{fig:groupspeed}. The occurrence of an (avoided) crossing is especially pronounced in animations of the group speed diagram when varying the wavenumber. Such animations are available as a supplement (see Supplementary materials).

For long and short wavelengths explicit limit expressions can be obtained from \cref{eq:groupspeed,eq:p_om,eq:p_k,eq:p_lam} and the phase speed limits in Table \ref{table:phasespeed}. The results are summarised in Tables \ref{table:groupspeed-shortwavelength} and \ref{table:groupspeed-longwavelength}. A similar summary was presented for a warm pair plasma in Ref. \onlinecite{Keppens2019_warmpair}. Substituting $\mu = 1$ and $w = v$ into our expressions we expect to recover the warm pair results. This is indeed the case except for the M and X mode. For these modes, the long wavelength group speed expressions in Table \ref{table:groupspeed-longwavelength} become identical because both $\omega_\mathrm{u}^2$ and $\omega_\mathrm{l}^2$ reduce to $1+E^2$. Their long wavelength group speeds were not the same in the warm pair case, however \cite{Keppens2019_warmpair}. Performing the warm pair substitutions this gives
\begin{equation}
\begin{aligned}
\frac{\partial\omega}{\partial\mathbf{k}} = \frac{k}{(1+E^2)^{3/2}} \bigg[ &\frac{(1-v^2-v^2E^2)+(-v^2E^2)}{2} \lambda \hat{\mathbf{b}} \\ &+ \frac{(v^2+v^2E^2)+(1+v^2E^2)}{2} \hat{\mathbf{n}} \bigg]
\end{aligned}
\end{equation}
which is the arithmetic average of the expressions in Ref. \onlinecite{Keppens2019_warmpair}. That this `discrepancy' exists between the warm pair plasma discussion in Ref. \onlinecite{Keppens2019_warmpair} and our current, general discussion for warm ion-electron plasmas may have to do with a lifted degeneracy as $w=v$ must be adopted. We note that the warm pair plasma dispersion relation turned out to always factorise in two branches (XFS versus OMA), whilst this exact factorisation is no longer possible in our current general treatment. 

\begin{table}[!htb]
\centering
\caption{Group speeds of all modes in the short wavelength limit (large $k$) assuming $E < E_{\mathrm{cr}}$. All labels refer to oblique angles ($0 < \lambda < 1$).}
\label{table:groupspeed-shortwavelength}
\begin{tabular}{l l}
\hline
Short wavelengths ($k \rightarrow \infty$) & \\[1ex]
\hline\\
$\displaystyle \left( \frac{\partial\omega}{\partial\mathbf{k}} \right)_\mathrm{X} = \hat{\mathbf{n}}$ & $\displaystyle \left( \frac{\partial\omega}{\partial\mathbf{k}} \right)_\mathrm{F} = w \hat{\mathbf{n}}$ \\[2ex]
$\displaystyle \left( \frac{\partial\omega}{\partial\mathbf{k}} \right)_\mathrm{O} = \hat{\mathbf{n}}$ & $\displaystyle \left( \frac{\partial\omega}{\partial\mathbf{k}} \right)_\mathrm{A} = \frac{E}{k} (\hat{\mathbf{b}} - \lambda \hat{\mathbf{n}})$ \\[2ex]
$\displaystyle \left( \frac{\partial\omega}{\partial\mathbf{k}} \right)_\mathrm{M} = v \hat{\mathbf{n}}$ & $\displaystyle \left( \frac{\partial\omega}{\partial\mathbf{k}} \right)_\mathrm{S} = \frac{I}{k} (\hat{\mathbf{b}} - \lambda \hat{\mathbf{n}})$ \\[3ex]
\hline
\end{tabular}
\end{table}

\begin{table*}[!htb]
\centering
\caption{Group speeds of all modes in the long wavelength limit (small $k$) assuming $E < E_{\mathrm{cr}}$. All labels refer to oblique angles ($0 < \lambda < 1$).}
\label{table:groupspeed-longwavelength}
\begin{tabular}{ l } 
\hline
Long wavelengths ($k \rightarrow 0$) \\[1ex]
\hline\\
$\begin{array}{lcl} \displaystyle \left( \frac{\partial\omega}{\partial\mathbf{k}} \right)_\mathrm{X} & = & \omega_\mathrm{u}^{-1} \left\{ 6\omega_\mathrm{u}^6 - 5\omega_\mathrm{u}^4 (E^2+I^2+3) + 4\omega_\mathrm{u}^2 \left[ E^2I^2 + (E+I)^2 + 3 \right] - 3(1+EI)^2 \right\}^{-1} \\ & & \bigg\{ \bigg( \omega_\mathrm{u}^6 (v^2+w^2+2) - \omega_\mathrm{u}^4 [ E^2w^2 + I^2v^2 + 2(E^2+I^2) + 2(v^2+w^2) + c_\mathrm{s}^2+4 ] \\ & & + \omega_\mathrm{u}^2 [E^2w^2+I^2v^2 + E^2+I^2 + EI(3+2EI+c_\mathrm{s}^2) + v^2+w^2+2c_\mathrm{s}^2+2] \\ & & - (1+EI)(EI+c_\mathrm{s}^2) \bigg) \hat{\mathbf{n}} - \lambda \bigg( \omega_\mathrm{u}^4 (E^2v^2+I^2w^2) \\ & & - \omega_\mathrm{u}^2 [E^2+I^2+EI((v^2+w^2)(3+EI)-3c_\mathrm{s}^2-1)] + EI(1+EI)(1+c_\mathrm{s}^2) \bigg) \hat{\mathbf{b}} \bigg\} \end{array}$ \\[2ex]
$\begin{array}{lcl} \displaystyle \left( \frac{\partial\omega}{\partial\mathbf{k}} \right)_\mathrm{O} & = & k \hat{\mathbf{n}} - \lambda k \big\{ EI \left[ 1+3c_\mathrm{s}^2 + (1+EI)(1+c_\mathrm{s}^2) - (3+EI)(v^2+w^2) \right] \\ & & + E^2 v^2 + I^2 w^2 -(E^2+I^2) \big\} \left[E^2I^2 - (E-I)^2\right]^{-1} \hat{\mathbf{b}} \end{array}$ \\[2ex]
$\begin{array}{lcl} \displaystyle \left( \frac{\partial\omega}{\partial\mathbf{k}} \right)_\mathrm{M} & = & \omega_\mathrm{l}^{-1} \left\{ 6\omega_\mathrm{l}^6 - 5\omega_\mathrm{l}^4 (E^2+I^2+3) + 4\omega_\mathrm{l}^2 \left[ E^2I^2 + (E+I)^2 + 3 \right] - 3(1+EI)^2 \right\}^{-1} \\ & & \bigg\{ \bigg( \omega_\mathrm{l}^6 (v^2+w^2+2) - \omega_\mathrm{l}^4 [ E^2w^2 + I^2v^2 + 2(E^2+I^2) + 2(v^2+w^2) + c_\mathrm{s}^2+4 ] \\ & & + \omega_\mathrm{l}^2 [E^2w^2+I^2v^2 + E^2+I^2 + EI(3+2EI+c_\mathrm{s}^2) + v^2+w^2+2c_\mathrm{s}^2+2] \\ & & - (1+EI)(EI+c_\mathrm{s}^2) \bigg) \hat{\mathbf{n}} - \lambda \bigg( \omega_\mathrm{l}^4 (E^2v^2+I^2w^2) \\ & & - \omega_\mathrm{l}^2 [E^2+I^2+EI((v^2+w^2)(3+EI)-3c_\mathrm{s}^2-1)] + EI(1+EI)(1+c_\mathrm{s}^2) \bigg) \hat{\mathbf{b}} \bigg\} \end{array}$ \\[2ex]
$\begin{array}{lcl} \displaystyle \left( \frac{\partial\omega}{\partial\mathbf{k}} \right)_\mathrm{F} & = & \displaystyle \frac{\left[ v_\mathrm{f}^2 (EI+c_\mathrm{s}^2) - \lambda^2 EI c_\mathrm{s}^2 \right] \hat{\mathbf{n}} + \lambda EI c_\mathrm{s}^2 (v_\mathrm{f}^2-1) \hat{\mathbf{b}}}{v_\mathrm{f} \left[ 2v_\mathrm{f}^2 (1+EI) -EI -c_\mathrm{s}^2 - \lambda^2 EI c_\mathrm{s}^2 \right]} \end{array}$ \\[2ex]
$\begin{array}{lcl} \displaystyle \left( \frac{\partial\omega}{\partial\mathbf{k}} \right)_\mathrm{A} & = & c_\mathrm{a} \hat{\mathbf{b}} \end{array}$ \\[2ex]
$\begin{array}{lcl} \displaystyle \left( \frac{\partial\omega}{\partial\mathbf{k}} \right)_\mathrm{S} & = & \displaystyle \frac{\left[ v_\mathrm{sl}^2 (EI+c_\mathrm{s}^2) - \lambda^2 EI c_\mathrm{s}^2 \right] \hat{\mathbf{n}} + \lambda EI c_\mathrm{s}^2 (v_\mathrm{sl}^2-1) \hat{\mathbf{b}}}{v_\mathrm{sl} \left[ 2v_\mathrm{sl}^2 (1+EI) -EI -c_\mathrm{s}^2 - \lambda^2 EI c_\mathrm{s}^2 \right]} \end{array}$ \\[3ex]
\hline
\end{tabular}
\end{table*}

\section{Comparisons}
The study of waves in two-fluid descriptions of ion-electron plasmas is discussed in many textbooks \cite{Stix1992, BoydSanderson2003, GurnettBhattacharjee2005, ThorneBlandford2017}, but is nowhere found to be as complete in its wave categorisations as our current general two-fluid discussion. In (up to recent) literature \cite{Stringer1963, Hameiri2005, Ishida2005, Damiano2009, Bellan2012, Zhao2015, Zhao2017}, approximate results are obtained or presented for special limits, like dispersion relations focusing only on high- versus low-frequency modes. With this general two-fluid framework, parallels can be drawn to some of these results in the literature and how they are retrieved or altered using our two-fluid approach. To this end, all bars on the frequency and wavenumber are made explicit again for ease of comparison to published results.

\subsection{Low-frequency waves}\label{sec:lfw}
In 1963, a dispersion relation for the low-frequency modes (S, A, and F) was derived in Ref. \onlinecite{Stringer1963} using a Hall MHD model, which was later recovered in Ref. \onlinecite{Bellan2012}. More recently, an ideal two-fluid model was used in Ref. \onlinecite{Zhao2014} to correct this Hall MHD equation. As a description of the low-frequency S, A, and F waves, this equation should be obtainable from our general dispersion relation (\ref{eq:disp-rel}) as an extension of the global, low-frequency MHD limit. In order to verify this, eq. $(\mathrm{A}17)$ in Ref. \onlinecite{Zhao2014} can be rewritten with our conventions and notation as
\begin{equation}\label{eq:lfw-disp}
\begin{aligned}
&( 1+\mu )^3 \Big\{ \bar{\omega}^6 (1+\bar{k})^2 - \bar{\omega}^4 \bar{k}^2 \big\{ \bar{k}^4 c_\mathrm{s}^2 \\ &+ \bar{k}^2 \left[ EI + 2c_\mathrm{s}^2 + \lambda^2 (E^2+I^2-EI) \right] + (1+\lambda^2) EI + c_\mathrm{s}^2 \big\} \\ &+ \bar{\omega}^2 \bar{k}^4 \lambda^2 \left\{ \bar{k}^2 c_\mathrm{s}^2 \left[ E^2 + I^2 \right] + EI (EI + 2c_\mathrm{s}^2) \right\} - \lambda^4 \bar{k}^6 E^2 I^2 c_\mathrm{s}^2 \Big\} \\
&\qquad = 0,
\end{aligned}
\end{equation}
using $Z = 1$ and the non-relativistic Alfv\'en speed expression $v_\mathrm{a}^2/c^2 \simeq EI$. From this equation the contributing terms in our much more general dispersion relation can be identified by counting powers of $\bar{\omega}^2$ and $\bar{k}^2$. The contributing terms are shown schematically in Table \ref{table:lfw-disp}. Comparing $\alpha_{03}$ to its equivalent in eq. (\ref{eq:lfw-disp}) reaffirms that a non-relativistic expression should be used for the Alfv\'en speed in Ref. \onlinecite{Zhao2014}. It is to be noted that this is already an important drawback of the reduced dispersion relation (\ref{eq:lfw-disp}), as our original equation was shown to be fully compatible with the relativistic expressions of the slow, Alfv\'en, and fast speeds.

\begin{table}[!htb]
\centering
\caption{Terms contributing to dispersion relation (\ref{eq:lfw-disp}) are highlighted in {\color{red} red}.}\label{table:lfw-disp}
	\begin{tabular}{ r | c c c c c } 
		& $1$ & $\bar{k}^2$ & $\bar{k}^4$ & $\bar{k}^6$ & $\bar{k}^8$ \\
		\hline
		$\bar{\omega}^{12}$ & $\alpha_{60}$ & & & & \\
		$\bar{\omega}^{10}$ & $\alpha_{50}$ & $\alpha_{51}$ & & & \\
		$\bar{\omega}^8$ & $\alpha_{40}$ & $\alpha_{41}$ & $\alpha_{42}$ & & \\
		$\bar{\omega}^6$ & {\color{red} $\alpha_{30}$} & {\color{red} $\alpha_{31}$} & {\color{red} $\alpha_{32}$} & $\alpha_{33}$ & \\
		$\bar{\omega}^4$ & & {\color{red} $\alpha_{21}$} & {\color{red} $\alpha_{22}$} & {\color{red} $\alpha_{23}$} & $\alpha_{24}$ \\
		$\bar{\omega}^2$ & & & {\color{red} $\alpha_{12}$} & {\color{red} $\alpha_{13}$} & $\alpha_{14}$ \\
		$1$ & & & & {\color{red} $\alpha_{03}$} & $\alpha_{04}$
	\end{tabular}
\end{table}

To go from the coefficients $\alpha_{mn}$ to the corresponding coefficients in eq. (\ref{eq:lfw-disp}), which we will call $\tilde{\alpha}_{mn}$, some further approximations are necessary (and an overall change of sign). First of all, note that that $v_\mathrm{a}^2/c^2 \simeq EI$ implies that $E$ and $I = \mu E$ are both $\mathcal{O}(c^{-1})$. Furthermore, $c_\mathrm{s}$, $v$, and $w$ are also $\mathcal{O}(c^{-1})$. To go from $\alpha_{mn}$ to $\tilde{\alpha}_{mn}$ can now be summarised as keeping only the terms of order $\mathcal{O}(c^{2(m-3)})$. This eliminates all $\alpha_{mn}$ with $m \geq 4$ as well as $\alpha_{33}$, $\alpha_{24}$, $\alpha_{14}$, and $\alpha_{04}$. For the remaining $\alpha_{mn}$, the highest order terms in $c$ are $\mathcal{O}(c^{2(m-3)})$. Effectively, this reduction gives the MHD limit extended with correction terms coming from the higher diagonals in Table \ref{table:lfw-disp}. Performing this reduction, we indeed recover eq. (\ref{eq:lfw-disp}) rather than the expressions in Ref. \onlinecite{Stringer1963} and \onlinecite{Bellan2012}. As pointed out in Ref. \onlinecite{Zhao2014}, this is due to the absence of some $\mathcal{O}(\mu)$ terms in Ref. \onlinecite{Stringer1963} and \onlinecite{Bellan2012}.

Whilst this is a meaningful reduction for non-relativistic regimes, there seems to be no obvious reason not to use the general dispersion relation for all 6 wave pairs in all regimes (unmagnetised to superstrongly magnetised, cold to warm), although an interest in purely the non-relativistic, low-frequency MHD waves for all wavelengths could use eq. (\ref{eq:lfw-disp}).

\subsection{Hall MHD}
The dispersion relation of (non-relativistic) Hall MHD reported on by Ref. \onlinecite{Hameiri2005} can also be recovered from our two-fluid dispersion relation. Rewriting their expression somewhat and using our conventions it becomes
\begin{equation}\label{eq:hmhd-disp}
\begin{aligned}
&\left(\frac{\omega}{k}\right)^6 - \left(\frac{\omega}{k}\right)^4 \left[ (1+\lambda^2)EIc^2 + v_\mathrm{s}^2 + \lambda_\mathrm{H} \lambda^2 EIc^2 \right] && \\ &+ \lambda^2 EIc^2 \,\left(\frac{\omega}{k}\right)^2 \left[ EIc^2 + 2v_\mathrm{s}^2 + \lambda_\mathrm{H} v_\mathrm{s}^2 \right] - \lambda^4 E^2I^2c^4 v_\mathrm{s}^2 &&= 0.
\end{aligned}
\end{equation}
Here, the Hall parameter $\lambda_\mathrm{H}$ is proportional to $\bar{k}^2$ as $\lambda_\mathrm{H} = \bar{k}^2 / \mu$. Since $\lambda_\mathrm{H} \sim k^2$, the prefactors of the Hall terms should be contained in $\alpha_{22}$ and $\alpha_{13}$. Indeed, multiplying the general dispersion relation (\ref{eq:disp-rel}) by $c^6/\bar{k}^6$ and taking the non-relativistic limit $c\rightarrow\infty$ and the $\mu = 0$ limit results without any further approximation in the non-relativistic Hall dispersion relation (\ref{eq:hmhd-disp}).

\subsection{Kinetic theory}\label{sec:kinetic}
In all textbooks on plasma physics and the literature, the two-fluid formalism is often used as a stepping stone for the much more involved treatment offered by kinetic theory \cite{GurnettBhattacharjee2005, ThorneBlandford2017}. Although it is well-known that some intricacies are inevitably lost in a two-fluid treatment, e.g. the velocity-space resonant aspect of Landau damping, a comparison between results from either approach can be meaningful to understand the possibilities and limits of the two-fluid treatment. Using Ref. \onlinecite{GurnettBhattacharjee2005} as guidance, a couple of results are compared here.

\subsubsection{Hot unmagnetised plasma}
The two-fluid analysis of waves in a hot unmagnetised plasma was presented in Sec. \ref{sec:unmagnetised}. All modes split with the exception of the F and M modes. Assuming $c_\mathrm{s} \simeq w$ and taking the immobile ion approximation where $w=0$, the quadratic branch describing these modes, given generally by
\begin{equation}\label{eq:kinetic-quadratic}
\bar{\omega}^{4} - \bar{\omega}^{2} \left[ 1 + \bar{k}^2 (v^2 + w^2) \right] + \bar{k}^2 (c_\mathrm{s}^2 + \bar{k}^2 v^{2} w^{2}) = 0,
\end{equation}
splits into a trivial mode (F mode) and $\omega^2 = \omega_\mathrm{p}^2 + v_\mathrm{e}^2 k^2$, the M mode or Langmuir wave. In kinetic theory, the Langmuir dispersion relation using a Maxwellian distribution for the thermal velocity gives \cite{GurnettBhattacharjee2005}
\begin{equation}
\omega^2 = \omega_\mathrm{p}^2 + 3 \left( \frac{p_\mathrm{e}}{n_\mathrm{e} m_\mathrm{e}} \right) k^2.
\end{equation}
From our definition of $v_\mathrm{e}$, this implies that $\gamma_\mathrm{e} = 3$. This means that the electrons influenced by the Langmuir wave move with only one degree of freedom. This information is not apparent from the two-fluid treatment in this paper as has been pointed out before in Ref. \onlinecite{GurnettBhattacharjee2005} using a less complete two-fluid approach.

A second result from the literature pertains to the phase speed of ion acoustic waves, which must relate to our F mode. Under the assumption that $w^2 \ll \bar{\omega}^2/\bar{k}^2 \ll v^2$, the quadratic branch (\ref{eq:kinetic-quadratic}) divided by $\bar{k}^4$ reduces to
\begin{equation}
\frac{\bar{\omega}^2}{\bar{k}^2} = \frac{c_\mathrm{s}^2}{1+\bar{k}^2v^2}
\end{equation}
after discarding the terms $\bar{\omega}^4/\bar{k}^4$, $-w^2 \bar{\omega}^2/\bar{k}^2$, and $v^2w^2$. If it is assumed that $Z T_\mathrm{e} \gg T_\mathrm{i}$ and $\mu \ll 1$, this result can be reformulated as
\begin{equation}
\frac{\omega^2}{k^2} = \frac{\mu v_\mathrm{e}^2}{1+k^2 v_\mathrm{e}^2 / \omega_\mathrm{pe}^2} = \frac{\mu v_\mathrm{e}^2}{1+k^2 \lambda_\mathrm{De}^2}
\end{equation}
where $\lambda_\mathrm{De}$ signifies the electron Debye length. This agrees with the result ($8.4.24$) of Ref. \onlinecite{GurnettBhattacharjee2005} where they use Maxwellian distributions for ions and electrons and consider the case $n_\mathrm{e} = n_\mathrm{i}$, which implies $Z = 1$ through charge neutrality. In that sense this result is more general, since it holds for any value of $Z$.

\subsubsection{Magnetised plasma}
At parallel propagation more known results from kinetic theory can be recovered in the two-fluid formalism. Consider once more the parallel quartic branch which can be factored as was done in eq. (\ref{eq:pl-quartic-factored}) (although we repeat here that this manner of writing the results mixes forward and backward wave pairs and is to be disfavored). These two factors can be reordered as
\begin{align}
&1 - \frac{c^2 k^2}{\omega^2} - \frac{\omega_\mathrm{pe}^2}{\omega (\omega + E)} - \frac{\omega_\mathrm{pi}^2}{\omega (\omega - I)} = 0 \label{eq:left-polarised} \\
\text{and}\quad &1 - \frac{c^2 k^2}{\omega^2} - \frac{\omega_\mathrm{pe}^2}{\omega (\omega - E)} - \frac{\omega_\mathrm{pi}^2}{\omega (\omega + I)} = 0 \label{eq:right-polarised}
\end{align}
which is how they appear in the kinetic theory literature \cite{GurnettBhattacharjee2005}. (Note though that in the literature the cyclotron frequencies are sometimes defined including the charge sign such that the electron cyclotron frequency is $-\Omega_\mathrm{e}$ rather than the $\Omega_\mathrm{e}$ used here.) The first factor then describes the left-hand polarised waves whilst the second one describes the right-hand polarised waves.

The polarisation properties of the various waves, as well as their electromagnetic versus electrostatic wave character, are invariably related to either purely parallel or purely perpendicular properties. In Ref. \onlinecite{Keppens2019_warmpair}, it was already discussed that the S, A, F, M, O, X labelling scheme necessarily abandons such terminology, since the waves simply show varying characteristics at small, intermediate to large wavelengths. This relates to a discussion of the eigenfunctions, in addition to the eigenfrequencies, which is left to future research.

Additionally, whistler waves \cite{Stix1992, Baumjohann1997, Bittencourt2004, GurnettBhattacharjee2005, Bellan2006, ThorneBlandford2017} are often discussed starting from the right-hand polarised factor (\ref{eq:right-polarised}) and discarding the constant term $1$ and the ion term. A more complete treatment of whistler waves is thus possible in the two-fluid framework, as was performed in Ref. \onlinecite{Huang2019}. However, if parallel crossings occur in the whistler frequency interval, the appearance of avoided crossings at oblique angles may lead to different whistler behaviour at exactly parallel and near-parallel propagation. This is left for future work.

\section{Conclusion}
Earlier discussions of plasma waves using a two-fluid treatment \cite{Keppens2019_coldpair, Keppens2019_coldei, Keppens2019_warmpair} were extended to the warm ion-electron plasma case. The previously introduced labelling scheme S, A, F, M, O, and X was shown to be unambiguous for warm ion-electron plasmas and preserves the wave mode frequency ordering at all oblique angles. Additionally, parallel and perpendicular propagation were confirmed as exceptional cases where the modes can cross. Hence, we argue against any wave type classification relying on parallel and perpendicular propagation like those present in many textbooks in favor of the S, A, F, M, O, X classification.

The mode crossings appearing at parallel and perpendicular propagation could all be identified, with up to 6 different possibilities in the number of crossings, depending on the plasma parameters. Approximate analytical expressions were given for all of them. Numerically, all crossings can be computed and frequency-wavenumber diagrams can be drawn for all angles. Complementary, the evolution of phase and group speed diagrams when varying the wavenumber further reveals the intricate structure and the unavoidable fact that all 6 wave modes, from MHD to Langmuir and electromagnetic wave modes, can be highly anisotropic in their phase and group speed behaviour.

With this complete two-fluid treatment a couple of results from the literature were recovered. Even though any ideal two-fluid methodology is unable to derive any damping effects, a selection of results from kinetic theory could also be recovered. Additionally, in the future one could also include collisional damping to compare the two-fluid and the kinetic dispersion relations in more detail.

The generality of our classification scheme is a major advantage over any earlier treatments that assume non-relativistic conditions, or adopt one or more species in the cold regime. All our results can be translated to diagrams focusing rather on the behaviour of the refractive index with wave and plasma parameters, or to equivalent descriptions that analyse all possible solutions at fixed frequency, such as done for obtaining the CMA diagrams \cite{GurnettBhattacharjee2005} of the cold plasma regime. Those latter diagrams actually render the wave interpretations even more ambiguous, since they confuse slow/fast and other wave modes, as explained in Ref. \onlinecite{Keppens2019_coldei}. We argue that the unambiguous S, A, F, M, O, X scheme is in that sense superior. The high frequency O and X waves that give well-known Faraday rotation effects \cite{Keppens2019_coldei} are now known for all orientations and plasma parameters.

\section*{Supplementary materials}
Animations of the phase and group speed diagrams whilst varying the wavenumber $k$ are available from \href{https://perswww.kuleuven.be/~u0016541/}{RK's homepage}.

\begin{acknowledgments}
This work is supported by funding from the European Research Council (ERC) under the European Unions Horizon 2020 research and innovation programme, Grant agreement No. 833251 PROMINENT ERC-ADG 2018. RK was additionally supported by a joint FWO-NSFC grant G0E9619N and by Internal funds KU Leuven, project C 14 / 19 / 089 TRACESpace.
\end{acknowledgments}

\section*{Data availability}
The data generated for this study are available from the corresponding author upon reasonable request.

\appendix
\section{Polynomial dispersion relation}\label{app:disprel}
The dispersion relation is a polynomial of the form
\begin{equation}
\sum\limits_{\substack{m,n \\ 3 \leq m+n \leq 6}} \alpha_{mn}\, \omega^{2m} k^{2n} = 0.
\end{equation}
To offer a complete overview, all coefficients $\alpha_{mn}$ are listed here in the  way they appear in Ref. \onlinecite{GoedbloedKeppensPoedts2019}.
\begin{widetext}
\begin{align}
\alpha_{60} &= 1, \\
\alpha_{50} &= -(3+E^2+I^2), \\
\alpha_{51} &= -(2+v^2+w^2), \\
\alpha_{40} &= 3+E^2+I^2+2EI+E^2I^2, \\
\alpha_{41} &= 4+2E^2+2I^2+(2+\lambda^2 E^2+I^2)v^2+(2+E^2+\lambda^2 I^2)w^2+c_\mathrm{s}^2, \\
\alpha_{42} &= 1+2v^2+2w^2+v^2w^2, \\
\alpha_{30} &= -(1+EI)^2, \\
\alpha_{31} &= -\bigg\{ 2(1+EI)^2 + (1+\lambda^2)(E^2+I^2-EI) + \left[ 1+I^2+\lambda^2 (3+EI)EI \right] v^2 \\ &\quad + \left[ 1+E^2+\lambda^2 (3+EI)EI \right] w^2 + \left[ 2+(1-3\lambda^2)EI \right] c_\mathrm{s}^2 \bigg\}, \\
\alpha_{32} &= -\bigg\{ 1+E^2+I^2+2(1+\lambda^2 E^2+I^2)v^2 + 2(1+E^2+\lambda^2 I^2)w^2 + 2c_\mathrm{s}^2 \\ &\quad + \left[ 2+\lambda^2(E^2+I^2) \right] v^2w^2 \bigg\}, \\
\alpha_{33} &= -(v^2+w^2+2v^2w^2), \\
\alpha_{21} &= (1+EI)(1+\lambda^2)EI + (1+EI)(1+\lambda^2 EI)c_\mathrm{s}^2, \\
\alpha_{22} &= (1+EI)EI + \lambda^2 (E^2+I^2-EI) + \left[ (1+\lambda^2)I^2 + 2\lambda^2 EI(2+EI) \right]v^2 \\ &\quad + \left[ (1+\lambda^2)E^2 + 2\lambda^2 EI(2+EI) \right]w^2 + \left[ 2+(1-5\lambda^2)EI \right]c_\mathrm{s}^2 + (1+\lambda^2 EI)^2 v^2w^2, \\
\alpha_{23} &= (I^2+\lambda^2 E^2)v^2 + (E^2+\lambda^2 I^2)w^2 + c_\mathrm{s}^2 + 2\left[ 1+\lambda^2 (E^2+I^2) \right] v^2w^2, \\
\alpha_{24} &= v^2w^2, \\
\alpha_{12} &= -\lambda^2 EI \left\{ EI + \left[ 2+(1+\lambda^2)EI \right] c_\mathrm{s}^2 \right\}, \\
\alpha_{13} &= -\lambda^2 \left\{ E^2I^2(v^2+w^2) + (E^2+I^2)c_\mathrm{s}^2 + 2EI (1+\lambda^2 EI) v^2w^2 \right\}, \\
\alpha_{14} &= -\lambda^2 (E^2+I^2) v^2w^2, \\
\alpha_{03} &= \lambda^4 E^2I^2 c_\mathrm{s}^2, \\
\alpha_{04} &= \lambda^4 E^2I^2 v^2w^2.
\end{align}

At perpendicular propagation, the dispersion relation factorises as
\begin{equation}\label{eq:disp-perp}
\begin{aligned}
&\omega^{4} \left(\omega^{2} - k^{2} - 1\right) \Big\{ \omega^{6} - \omega^{4} \left[ 2 + E^{2} + I^{2} + k^{2} \left( 1 + v^{2} + w^{2} \right)\right] && \\ &+ \omega^{2} \left[ (1 + EI)^2 + k^{2} \left( 1 + E^2 + I^2 + E^2 w^2 + I^2 v^2 + v^2 + w^2 + c_\mathrm{s}^2 \right) + k^{4} \left(v^{2} w^{2} + v^{2} + w^{2}\right) \right] && \\ &- k^2 \left[ (1+EI)(EI+c_\mathrm{s}^2) + k^2 \left(E^{2} w^{2} + I^2 v^{2} + v^{2} w^{2} + c_\mathrm{s}^2\right) + k^4 v^{2} w^{2} \right] \Big\} &&= 0.
\end{aligned}
\end{equation}
\end{widetext}

\section{Test cases for parallel propagation}\label{app:testcases}
In Sec. \ref{sec:dispdiagram-pl} multiple analytical approximations were offered for various crossings in different regimes. These expressions were obtained by taking single term approximations of the quadratic branch for long and short wavelengths. Naturally, it should be checked how accurate the final expressions are. In order to test this accuracy, the analytical results were compared with numerical solutions. Since the crossings and the accuracy vary based on the regime, test cases were used for each regime. For the $1 < E < E_{\mathrm{cr}}$ regimes two test cases were used since some approximations were less accurate for values of $E$ closer to $1$. All test cases and their crossings are described in Table \ref{table:testcases}.

\begin{table*}[!htb]
\caption{Comparison of numerical crossings and analytical approximations. Each regime is represented by a test case (or two) specifying $E$, $v$, and $w$. All cases use $\mu = 1/1836$. Numerical solutions of crossings are given alongside their analytical approximations.}\label{table:testcases} 
\begin{tabular}{| l | l | l | l | l |}
\hline
Regime & Parameters & Label & Numerical crossings & Analytical approximations \\
\hline
$\begin{aligned} &E < 1, \\ &c_\mathrm{s} < c_\mathrm{a} \end{aligned}$ & $\begin{aligned} E &= 0.5 \\ v &= 0.002 \\ w &= 0.0005 \end{aligned}$ & $\begin{aligned} &\text{SA} \\ &\text{SF} \\ &\text{MO} \end{aligned}$ & $\begin{aligned} &\texttt{(2.71827346e-04, 5.41448592e-01)} \\ &\texttt{(4.99999496e-01, 9.95941087e+02)} \\ &\texttt{(1.00000067, 5.77194e-01)} \end{aligned}$ & $\begin{aligned} &\texttt{(2.71827346e-04, 5.41448590e-01)} \\ &\texttt{(4.99999496e-01, 9.95941087e+02)} \\ &\texttt{(1, 5.77193-01)} \end{aligned}$\\
\hline
$\begin{aligned} &E < 1, \\ &c_\mathrm{s} > c_\mathrm{a} \\ &\text{(Fig. \ref{fig:pl_Evar}a)} \end{aligned}$ & $\begin{aligned} E &= 0.5 \\ v &= 0.1 \\ w &= 0.05 \end{aligned}$ & $\begin{aligned} &\text{AF} \\ &\text{AF} \\ &\text{MO} \end{aligned}$ & $\begin{aligned} &\texttt{(0.00479704 0.0958626)} \\ &\texttt{(0.49493417 9.89402523)} \\ &\texttt{(1.0016786  0.57977526)} \end{aligned}$ & $\begin{aligned} &\texttt{(0.00479704, 0.0958626)} \\ &\texttt{(0.49493063, 9.89053971)} \\ &\texttt{(1, 0.57719297)} \end{aligned}$ \\
\hline
$\begin{aligned} &1 < E < E_{\mathrm{cr}}, \\ &c_\mathrm{s} < c_\mathrm{a} \end{aligned}$ & $\begin{aligned} E &= 2 \\ v &= 0.01 \\ w &= 0.005 \end{aligned}$ & $\begin{aligned} &\text{SA} \\ &\text{SF} \\ &\text{MO} \\ &\text{FM} \\ &\text{FM} \end{aligned}$ & $\begin{aligned} &\texttt{(0.00107681, 0.21518649)} \\ &\texttt{(1.99998749, 399.67127926)} \\ &\texttt{(1.0000333, 0.81631022)} \\ &\texttt{(1.00009992, 1.41397011)} \\ &\texttt{(1.99995002, 199.9950021)} \end{aligned}$ & $\begin{aligned} &\texttt{(0.00107681, 0.21518649)} \\ &\texttt{(1.99998749, 399.67127926)} \\ &\texttt{(1, 0.81627395)} \\ &\texttt{(1, 1.41382879)} \\ &\texttt{(1.99995002, 199.9950021)} \end{aligned}$ \\
\hline
$\begin{aligned} &1 < E < E_{\mathrm{cr}}, \\ &c_\mathrm{s} < c_\mathrm{a} \\ &\text{(Fig. \ref{fig:pl_Evar}b)} \end{aligned}$ & $\begin{aligned} E &= 10 \\ v &= 0.1 \\ w &= 0.05 \end{aligned}$ & $\begin{aligned} &\text{SA} \\ &\text{SF} \\ &\text{MO} \\ &\text{FM} \\ &\text{FM} \end{aligned}$ & $\begin{aligned} &\texttt{(0.00519572, 0.10382959)} \\ &\texttt{(9.99974909, 199.83187597)} \\ &\texttt{(1.0045746, 0.95780243)} \\ &\texttt{(1.00559418, 1.05944303)} \\ &\texttt{(9.99898056, 99.51358858)} \end{aligned}$ & $\begin{aligned} &\texttt{(0.00519572, 0.10382958)} \\ &\texttt{(9.99974909, 199.83187597)} \\ &\texttt{(1, 0.95320147)} \\ &\texttt{(1, 1.05380701)} \\ &\texttt{(9.99899035, 99.98990347)} \end{aligned}$ \\
\hline
$\begin{aligned} &1 < E < E_{\mathrm{cr}}, \\ &c_\mathrm{s} > c_\mathrm{a} \end{aligned}$ & $\begin{aligned} E &= 2 \\ v &= 0.3 \\ w &= 0.25 \end{aligned}$ & $\begin{aligned} &\text{AF} \\ &\text{AF} \\ &\text{MO} \\ &\text{AM} \\ &\text{AM} \end{aligned}$ & $\begin{aligned} &\texttt{(0.03280859, 0.13121866)} \\ &\texttt{(1.96610763, 7.86408174)} \\ &\texttt{(1.03208004, 0.85114203)} \\ &\texttt{(1.10491084, 1.56653326)} \\ &\texttt{(1.92716371, 5.49161664)} \end{aligned}$ & $\begin{aligned} &\texttt{(0.03280860, 0.13121867)} \\ &\texttt{(1.96610208, 7.86346664)} \\ &\texttt{(1, 0.81627395)} \\ &\texttt{(1, 1.41382879)} \\ &\texttt{(1.94929139, 6.49763796)} \end{aligned}$ \\
\hline
$\begin{aligned} &1 < E < E_{\mathrm{cr}}, \\ &c_\mathrm{s} > c_\mathrm{a} \end{aligned}$ & $\begin{aligned} E &= 10 \\ v &= 0.3 \\ w &= 0.25 \end{aligned}$ & $\begin{aligned} &\text{AF} \\ &\text{AF} \\ &\text{MO} \\ &\text{AM} \\ &\text{AM} \end{aligned}$ & $\begin{aligned} &\texttt{(0.00122256, 0.00488966)} \\ &\texttt{(9.99333248, 39.9732234)} \\ &\texttt{(1.04378574, 0.99723593)} \\ &\texttt{(1.05378906, 1.107997)} \\ &\texttt{(9.98999528, 33.132819)} \end{aligned}$ & $\begin{aligned} &\texttt{(0.00122256, 0.00488966)} \\ &\texttt{(9.99333081, 39.9685369)} \\ &\texttt{(1, 0.95320147)} \\ &\texttt{(1, 1.05380701)} \\ &\texttt{(9.99010549, 33.300352)} \end{aligned}$ \\
\hline
$\begin{aligned} &E_{\mathrm{cr}} < E < 1/\mu, \\ &c_\mathrm{s} < c_\mathrm{a} \end{aligned}$ & $\begin{aligned} E &= 1835.5 \\ v &= 0.1 \\ w &= 0.05 \end{aligned}$ & $\begin{aligned} &\text{SA} \\ &\text{SF} \\ &\text{FM} \\ &\text{FM} \end{aligned}$ & $\begin{aligned} &\texttt{(0.9997263, 19.99046107)} \\ &\texttt{(1835.49999863, 36680.06113394)} \\ &\texttt{(1.00503714, 1.00517377)} \\ &\texttt{(1835.4999945, 18354.99722243)} \end{aligned}$ & $\begin{aligned} &\texttt{(0.9997263, 19.97821950)} \\ &\texttt{(1835.49999863, 36680.06113394)} \\ &\texttt{(1, 1.00013629)} \\ &\texttt{(1835.4999945, 18354.99994500)} \end{aligned}$ \\
\hline
$\begin{aligned} &E_{\mathrm{cr}} < E < 1/\mu, \\ &c_\mathrm{s} > c_\mathrm{a} \end{aligned}$ & \multicolumn{4}{c|}{Unphysical} \\
\hline
$\begin{aligned} &E > 1/\mu, \\ &c_\mathrm{s} < c_\mathrm{a} \\ &\text{(Fig. \ref{fig:pl_Evar}c)} \end{aligned}$ & $\begin{aligned} E &= 2500 \\ v &= 0.1 \\ w &= 0.05 \end{aligned}$ & $\begin{aligned} &\text{SA} \\ &\text{SF} \\ &\text{AM} \\ &\text{FM} \\ &\text{FM} \\ &\text{AM} \end{aligned}$ & $\begin{aligned} &\texttt{(1.36165477, 27.22974249)} \\ &\texttt{(2499.999999, 49959.22248234)} \\ &\texttt{(1.00504145, 1.00560472)} \\ &\texttt{(1.00503662, 1.00512158)} \\ &\texttt{(2499.99999596, 24999.99995962)} \\ &\texttt{(1.36165165, 13.47752584)} \end{aligned}$ & $\begin{aligned} &\texttt{(1.36165477, 27.21088544)} \\ &\texttt{(2499.999999, 49959.22248234)} \\ &\texttt{(1, 1.00055264)} \\ &\texttt{(1, 1.00008472)} \\ &\texttt{(2499.99999596, 24999.99995962)} \\ &\texttt{(1.36165173, 13.61651735)} \end{aligned}$ \\
\hline
$\begin{aligned} &E > 1/\mu, \\ &c_\mathrm{s} > c_\mathrm{a} \end{aligned}$ & \multicolumn{4}{c|}{Unphysical} \\
\hline
\end{tabular}
\end{table*}

\section{Group speed polynomials}\label{app:group_polynom}
As noted in Section \ref{sec:group-speed}, the group speed expressions are of the form
\begin{equation}
\frac{\partial\omega}{\partial\mathbf{k}} = -(v_{\mathrm{ph}} P_\omega)^{-1}\left[ P_k \hat{\mathbf{n}} + \frac{\lambda P_\lambda}{k^2} \left( \hat{\mathbf{b}} - \lambda\hat{\mathbf{n}} \right) \right]
\end{equation}
where the polynomials are given by the expressions
\begin{align}
P_\omega &= \sum\limits_{\substack{1 \leq m \\ 0\leq n \\ 3 \leq m+n \leq 6}} m \alpha_{mn}\, \omega^{2(m-1)} k^{2n}, \\
P_k &= \sum\limits_{\substack{0\leq m \\ 1 \leq n \\ 3 \leq m+n \leq 6}} n \alpha_{mn}\, \omega^{2m} k^{2(n-1)}, \\
\text{and}\quad P_\lambda &= \sum\limits_{\substack{0 \leq m,n \\ 3 \leq m+n \leq 6}} \frac{\partial\alpha_{mn}}{\partial\lambda^2} \, \omega^{2m} k^{2n}.
\end{align}

\begin{table*}[!htb]
\centering
\caption{In the left table, the terms contributing to $P_\omega$ are highlighted in {\color{NavyBlue} blue}. In the middle table, those contributing to $P_k$ are indicated in {\color{Mulberry} purple}. In the right table, the terms with a contribution to $P_\lambda$ are marked in {\color{red} red}.}
\label{table:group-polynom}
	\begin{tabular}{ r | c c c c c } 
		& $1$ & $k^2$ & $k^4$ & $k^6$ & $k^8$ \\
		\hline
		$\omega^{12}$ & {\color{NavyBlue} $\alpha_{60}$} & & & & \\
		$\omega^{10}$ & {\color{NavyBlue} $\alpha_{50}$} & {\color{NavyBlue} $\alpha_{51}$} & & & \\
		$\omega^8$ & {\color{NavyBlue} $\alpha_{40}$} & {\color{NavyBlue} $\alpha_{41}$} & {\color{NavyBlue} $\alpha_{42}$} & & \\
		$\omega^6$ & {\color{NavyBlue} $\alpha_{30}$} & {\color{NavyBlue} $\alpha_{31}$} & {\color{NavyBlue} $\alpha_{32}$} & {\color{NavyBlue} $\alpha_{33}$} & \\
		$\omega^4$ & & {\color{NavyBlue} $\alpha_{21}$} & {\color{NavyBlue} $\alpha_{22}$} & {\color{NavyBlue} $\alpha_{23}$} & {\color{NavyBlue} $\alpha_{24}$} \\
		$\omega^2$ & & & {\color{NavyBlue} $\alpha_{12}$} & {\color{NavyBlue} $\alpha_{13}$} & {\color{NavyBlue} $\alpha_{14}$} \\
		$1$ & & & & $\alpha_{03}$ & $\alpha_{04}$
	\end{tabular}
	\qquad
	\begin{tabular}{ r | c c c c c } 
		& $1$ & $k^2$ & $k^4$ & $k^6$ & $k^8$ \\
		\hline
		$\omega^{12}$ & $\alpha_{60}$ & & & & \\
		$\omega^{10}$ & $\alpha_{50}$ & {\color{Mulberry} $\alpha_{51}$} & & & \\
		$\omega^8$ & $\alpha_{40}$ & {\color{Mulberry} $\alpha_{41}$} & {\color{Mulberry} $\alpha_{42}$} & & \\
		$\omega^6$ & $\alpha_{30}$ & {\color{Mulberry} $\alpha_{31}$} & {\color{Mulberry} $\alpha_{32}$} & {\color{Mulberry} $\alpha_{33}$} & \\
		$\omega^4$ & & {\color{Mulberry} $\alpha_{21}$} & {\color{Mulberry} $\alpha_{22}$} & {\color{Mulberry} $\alpha_{23}$} & {\color{Mulberry} $\alpha_{24}$} \\
		$\omega^2$ & & & {\color{Mulberry} $\alpha_{12}$} & {\color{Mulberry} $\alpha_{13}$} & {\color{Mulberry} $\alpha_{14}$} \\
		$1$ & & & & {\color{Mulberry} $\alpha_{03}$} & {\color{Mulberry} $\alpha_{04}$}
	\end{tabular}
	\qquad
	\begin{tabular}{ r | c c c c c } 
		& $1$ & $k^2$ & $k^4$ & $k^6$ & $k^8$ \\
		\hline
		$\omega^{12}$ & $\alpha_{60}$ & & & & \\
		$\omega^{10}$ & $\alpha_{50}$ & $\alpha_{51}$ & & & \\
		$\omega^8$ & $\alpha_{40}$ & {\color{red} $\alpha_{41}$} & $\alpha_{42}$ & & \\
		$\omega^6$ & $\alpha_{30}$ & {\color{red} $\alpha_{31}$} & {\color{red} $\alpha_{32}$} & $\alpha_{33}$ & \\
		$\omega^4$ & & {\color{red} $\alpha_{21}$} & {\color{red} $\alpha_{22}$} & {\color{red} $\alpha_{23}$} & $\alpha_{24}$ \\
		$\omega^2$ & & & {\color{red} $\alpha_{12}$} & {\color{red} $\alpha_{13}$} & {\color{red} $\alpha_{14}$} \\
		$1$ & & & & {\color{red} $\alpha_{03}$} & {\color{red} $\alpha_{04}$}
	\end{tabular}
\end{table*}

Which terms contribute to each polynomial is visualised in Table \ref{table:group-polynom}. The factors $\alpha_{mn}$ that appear in the expressions for $P_\omega$ and $P_k$ were already stated in App. \ref{app:disprel}. However, the expression for $P_\lambda$ contains factors of the form $\partial\alpha_{mn} / \partial\lambda^2$. For the sake of completeness, these derivatives are
\begin{align}
\frac{\partial\alpha_{41}}{\partial\lambda^2} =\quad &E^2 v^2 + I^2 w^2 \\
\frac{\partial\alpha_{31}}{\partial\lambda^2} =\quad &-\left\{ E^2 + I^2 - EI + EI (3+EI) (v^2+w^2) - 3EI c_\mathrm{s}^2 \right\} \\
\frac{\partial\alpha_{32}}{\partial\lambda^2} =\quad &-\left\{ 2 (E^2 v^2 + I^2 w^2) + (E^2 + I^2) v^2 w^2 \right\} \\
\frac{\partial\alpha_{21}}{\partial\lambda^2} =\quad &EI (1+EI) (1+c_\mathrm{s}^2) \\
\frac{\partial\alpha_{22}}{\partial\lambda^2} =\quad &E^2 + I^2 - EI + I^2 v^2 + E^2 w^2 + 2EI (2+EI) (v^2 + w^2) \nonumber \\ &- 5EI c_\mathrm{s}^2 + 2EI (1+\lambda^2 EI) v^2 w^2 \\
\frac{\partial\alpha_{23}}{\partial\lambda^2} =\quad &E^2 v^2 + I^2 w^2 + 2 (E^2 + I^2) v^2 w^2 \\
\frac{\partial\alpha_{12}}{\partial\lambda^2} =\quad &-EI \left\{ EI + \left[ 2 + (1+2\lambda^2) EI \right] c_\mathrm{s}^2 \right\} \\
\frac{\partial\alpha_{13}}{\partial\lambda^2} =\quad &- \{ E^2 I^2 (v^2 + w^2) + (E^2 + I^2) c_\mathrm{s}^2 \nonumber \\ &+ 2EI (1+2\lambda^2 EI) v^2 w^2 \} \\
\frac{\partial\alpha_{14}}{\partial\lambda^2} =\quad &-(E^2 + I^2) v^2 w^2 \\
\frac{\partial\alpha_{03}}{\partial\lambda^2} =\quad &2 \lambda^2 E^2 I^2 c_\mathrm{s}^2 \\
\frac{\partial\alpha_{04}}{\partial\lambda^2} =\quad &2 \lambda^2 E^2 I^2 v^2 w^2
\end{align}

\bibliography{bibliography.bib}
\end{document}